\documentclass{aa}  

\usepackage{natbib}
\usepackage{graphicx}
\usepackage{txfonts}
\usepackage{xcolor}
\usepackage[colorlinks=true,citecolor=blue]{hyperref}
\usepackage{float}
\usepackage{twoopt}
\makeatletter
  \renewcommand*\aa@pageof{, page \thepage{} of \pageref*{LastPage}}
  \newcommandtwoopt{\citeads}[3][][]{\href{http://adsabs.harvard.edu/abs/#3}%
    {\def\hyper@linkstart##1##2{}%
     \let\hyper@linkend\@empty\citealp[#1][#2]{#3}}}
  \newcommandtwoopt{\citepads}[3][][]{\href{http://adsabs.harvard.edu/abs/#3}%
    {\def\hyper@linkstart##1##2{}%
     \let\hyper@linkend\@empty\citep[#1][#2]{#3}}}
  \newcommandtwoopt{\citetads}[3][][]{\href{http://adsabs.harvard.edu/abs/#3}%
    {\def\hyper@linkstart##1##2{}%
     \let\hyper@linkend\@empty\citet[#1][#2]{#3}}}
  \newcommandtwoopt{\citeyearads}[3][][]%
    {\href{http://adsabs.harvard.edu/abs/#3}
    {\def\hyper@linkstart##1##2{}%
     \let\hyper@linkend\@empty\citeyear[#1][#2]{#3}}}
\makeatother

\usepackage{color}
\newcommand{\Msun}{{M}_{\odot}} 

\defcitealias{ChruslinskaNelemans19}{ChN19}

\begin{document}

   \title{The effect of the environment-dependent IMF on the formation and
   metallicities of stars over the cosmic history}

\titlerunning{Non-universal IMF and the metallicity of stars}

   \author{M. Chru{\'s}li{\'n}ska
          \inst{1}
          \and
          T. Je{\v{r}}{\'a}bkov{\'a}\inst{2,3,4,5,6}
          \and
          G. Nelemans\inst{1,7,8}
          \and
          Z. Yan\inst{2,3}
          }

   \institute{Department of Astrophysics/IMAPP,
   Radboud University, P O Box 9010, 
   NL-6500 GL Nijmegen, The Netherlands\\
   \email{m.chruslinska@astro.ru.nl}
    \and 
  Helmholtz-Institut f{\"u}r Strahlen- und Kernphysik (HISKP), Universität Bonn, Nussallee 14–16, 53115 Bonn, Germany
             \\ \email{yan@astro.uni-bonn.de}  
         \and
             Charles University in Prague, Faculty of Mathematics and Physics, Astronomical Institute, V Hole{\v s}ovi{\v c}k{\'a}ch 2, CZ-180 00 Praha 8, Czech Republic
        \and 
            Astronomical Institute, Czech Academy of Sciences, Fri\v{c}ova 298, 25165, Ond\v{r}ejov, Czech Republic
        \and 
            Instituto de Astrof{\'i}sica de Canarias, E-38205 La Laguna, Tenerife, Spain
        \and
            GRANTECAN, Cuesta de San Jose s/n, 38712 Brena Baja, La Palma, Spain\\
            \email{tereza.jerabkova@gtc.iac.es}
         \and
            Institute for Astronomy, KU Leuven, Celestijnenlaan 200D, 3001 Leuven, Belgium
          \and
            SRON, Netherlands Institute for Space Research, Sorbonnelaan 2, NL-3584 CA Utrecht, the Netherlands
             }

   \date{Received February 07, 2020; accepted February 22, 2020}

 
  \abstract
  {
Recent observational and theoretical studies indicate that the stellar initial mass function (IMF) varies systematically with the environment (star formation rate - SFR, metallicity). Although the exact dependence of the IMF on those properties is likely to
change with improving observational constraints, the reported trend in the shape of the IMF appears robust. We present the first study aiming to evaluate the effect of the IMF variations on the measured cosmic SFR density (SFRD) as a function of metallicity and redshift, $f_{\rm SFR}$(Z,z).
We also study the expected number and metallicity of white dwarf, neutron star and black hole progenitors under different IMF assumptions. Applying the empirically driven IMF variations described by the integrated galactic IMF (IGIMF) theory, we correct $f_{\rm SFR}$(Z,z) obtained by Chruslinska \& Nelemans (2019) and find lower SFRD at high redshifts as well as a higher fraction of metal-poor stars being formed. In the local Universe, our calculation applying the IGIMF theory suggests more white dwarf and  neutron star progenitors in comparison with the universal IMF scenario,
while the number of black hole progenitors remains unaffected.
}

   \maketitle

\section{Introduction}

The distribution of the cosmic star formation rate density (SFRD) at different metallicities and redshifts, $f_{\rm SFR}$(Z,z), is a necessary ingredient e.g. to estimate the rate of occurrence of any stellar or binary evolution related phenomena. Among those, double compact object mergers and long gamma ray bursts are particularly sensitive to metallicity and their inferred rates, as well as the properties of their progenitor systems can vary significantly depending on the assumed $f_{\rm SFR}$(Z,z) \citep[e.g.][]{Chruslinska19, Neijssel19}. Understanding the uncertainty of this distribution may be crucial to correctly compare the theoretical estimates of the properties of those objects with observations and learn about the physics behind the formation and evolution of their progenitors.
\\
Recently, Chruslinska \& Nelemans (2019; hereafter \citetalias{ChruslinskaNelemans19}) endeavoured to find the observation-based $f_{\rm SFR}$(Z,z) distribution and to evaluate its uncertainty due to the currently unresolved issues involved in the observations of star forming galaxies. 
These include the unknown source of differences between the metallicity estimates obtained with different calibrations \citep[e.g.][]{MaiolinoMannucci19}, the poorly constrained low mass end of the galaxy stellar mass function \citep[e.g.][]{Conselice16} and the shape of the high mass end of the star formation -- mass relation (SFMR; some studies report the flattening of the high-mass part of the relation e.g. \citealt{Whitaker14,Lee15,Tomczak16}, while other find no evidence for such a flattening, e.g. \citealt{Speagle14,Pearson18})
\\
\citetalias{ChruslinskaNelemans19} assume an universal, \citet{Kroupa01} initial mass function (IMF) but remark that the systematic variations in the IMF with the star formation rate (SFR) or/and metallicity (Z) may have a significant and non-straightforward impact on their results. However, the uncertainty of $f_{\rm SFR}$(Z,z) coming from this assumption could not be evaluated.
\\
In fact, 
systematic variations of the IMF 
are expected on theoretical grounds \citep{Larson1998, Adams_Fatuzzo1996, Adams_Laughlin1996, Dib2007, Papadopoulos2010}
 and have now been reported in many observational studies \citep[see e.g. reviews by][]{Kroupa13,Hopkins18}. There is observational indication of a top-heavy IMF in low metallicity and high gas density or high star-formation rate systems \citep[e.g.][]{Matteucci1994,Dabringhausen2009,Papadopoulos2010,Marks2012TH, Dabringhausen2012, Zhang18, Brown2019, Kalari2018, Schneider18}, top-light IMF in low star-formation rate systems \citep[e.g.][]{Lee2009,Meurer2009,Watts2018}, bottom-heavy IMF in metal-rich environments \citep[e.g.][]{Kroupa2002, Marks2012TH, Conroy2017, Martin-Navarro15}. A time-variable IMF changing with evolving environmental properties is needed in order to explain all observational constraints (such as low-mass spectral features and chemical abundances, such as [Mg/Fe], e.g \citealt{Vazdekis1997,Weidner13a, Ferraras2015,Gargiulo2015,Fontanot2017,Jerabkova18}).
 A promising method to model the IMF variations, building on the integrated galactic IMF theory (IGIMF) put forward by \citet{KroupaWeidner03}, has been recently developed by \citet{Yan17,Jerabkova18,Yan2019}. 
\\
In this contribution we aim to evaluate the possible impact of the SFR- and metallicity-dependent IMF on the $f_{\rm SFR}$(Z,z). 
We combine the method described in \citetalias{ChruslinskaNelemans19}
with the IGIMF, using the SFR and metallicity dependent IMF model from
\citet{Jerabkova18}. Our approach is summarised in Section \ref{sec: method}. Our results are shown in Sec. \ref{sec: results}; we focus on the differences in the $f_{\rm SFR}$(Z,z) obtained in this study with respect to the
distributions found in \citetalias{ChruslinskaNelemans19}. 
\\
In general, the effect of the transition from an universal to non-universal IMF will be different for stars forming in different mass ranges.
Therefore, we also discuss the implications of the employed non-universal IMF model for the expected number and metallicity of stars forming in different mass ranges across  the  cosmic  history.
This is important for the interpretation (in case of observations) or estimation (in case of theoretical studies) of the properties of populations of (objects composed of) stars and their remnants.
It also affects the expected rates of occurrence of various stellar evolution-related phenomena,
because the formation efficiency of many transients of stellar/binary origin (e.g. double black hole mergers, long gamma ray bursts) depends on metallicity.
Our results can provide a guidance on whether the properties of stellar populations inferred under the assumption of the
 universal IMF are likely to under/over-predict the estimated quantities and on the order of magnitude of
this effect.
The results of our calculations are available at \url{https://ftp.science.ru.nl/astro/mchruslinska/}.
\\
Where appropriate we adopt a standard flat cosmology with $\Omega_{M}$=0.3,$\Omega_{\Lambda}$=0.7, and $H_{0}$=70 km s$^{-1}$ Mpc$^{-1}$.
All mentions of the universal or canonical IMF throughout this study 
refer to the \citet[hereafter K01]{Kroupa01} IMF between 0.08 to 120 solar masses ($M_{\odot}$).

\section{Method}\label{sec: method}

To estimate $f_{\rm SFR}$(Z,z) under the assumption of an evolving, SFR and metallicity dependent IMF we combine the methods detailed in \citetalias{ChruslinskaNelemans19} and \citet{Jerabkova18}. Below we focus on the information relevant for this study and refer the reader to the original papers for further details.

\subsection{The distribution of the cosmic SFR at different metallicities and redshift}

\citetalias{ChruslinskaNelemans19} construct the $f_{\rm SFR}$(Z,z) based on the empirical scaling relations for star forming galaxies combined over a wide range of redshifts and stellar masses of galaxies ($M_{\rm gal,\ast}$). The observed galaxy stellar mass function (GSMF) of star forming galaxies is used to obtain the number density of objects of different masses. The metallicity (oxygen abundance) is then assigned to each mass with the use of the gas-based  mass--metallicity relation (MZR). The star-formation--mass relation (SFMR) sets the contribution of galaxies of different masses (metallicities) to the total SFRD at a certain redshift. Combining them, one can obtain the $f_{\rm SFR}$(Z,z). The intrinsic scatter present in the relations, the internal distribution of metallicities in the star-forming gas within galaxies and 
the so-called fundamental metallicity relation \citep[i.e. the empirical relation between $M_{\rm gal,\ast}$, metallicity and SFR of galaxies, which shows that galaxies of the same $M_{\rm gal,\ast}$ that have higher than average SFR at the same time have lower than average metallicities, e.g. ][]{Mannucci10} are taken into account.
The above mentioned empirical relations, as well as the results
shown in \citetalias{ChruslinskaNelemans19} obtained with the use of those relations assume the universal IMF.

\subsection{The environment dependent galaxy-wide IMF}
\begin{figure}
    \centering
    \includegraphics[width=\hsize]{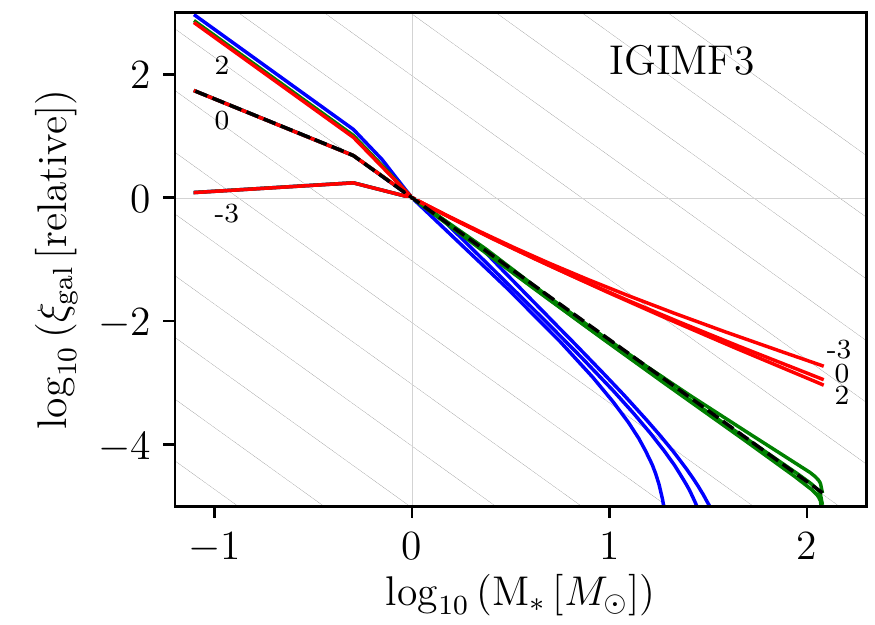}
    \caption{Galaxy-wide IMFs computed using IGIMF3 as a function of stellar mass plotted for several values of SFRs ($10^{-3}$ -- blue, $1.0$ -- green, $10^{3}$ -- red $M_{\odot}$/yr) and [Fe/H] (-3, 0 2). The galaxy-wide IMFs are normalised by their values at 1 $M_{\odot}$ to show the slope changes. The universal K01 IMF is plotted as black dashed line.
    }
    \label{fig:IGIMF_3}
\end{figure}

To model the non-universal, environment dependent IMF we rely on the
description provided by the IGIMF theory \citep{Weidner13,Yan17,Jerabkova18}. 
\\
The IGIMF uses empirically derived prescriptions for the formation of stars on pc/embedded cluster scales to build-up galaxy-wide stellar populations.
Those prescriptions are based on a few key inferences from the observations of the local star forming environments: (i) all stars form in embedded star clusters
\citep{Jerabkova18, lada2003,Kroupa2005, Megeath2016, Joncour2018}; (ii) the IMF of each embedded star cluster varies with the gas density and metallicity of the star forming material in the cluster 
\citep[i.e. the becomes more top heavy in high gas density and low metallicity environments e.g.][]{Kroupa02,Dabringhausen2009,Marks12}
and is well approximated by a broken power law;\footnote{We note that the IMF differs significantly from the canonical IMF mainly for extreme densities and very low metallicities.} (iii) initial masses of star clusters are distributed according to a single power law, the slope and the upper mass limit of the distribution depend on the SFR of a galaxy.
\\
The star clusters with the same mass, metallicity and density are assumed to lead to the
same cluster IMF independent of their redshift.
The galaxy wide IMF and stellar population is then obtained by summing the contributions from the local galactic star-forming regions in a given time interval.
\\
The IGIMF theory is consistent with the Milky Way stellar population /galaxy-wide IMF.
This is an important consistency check, since it is not clear a priori that adding all the IMFs in all embedded clusters would yield a galaxy-wide IMF which agrees with the observational constraints \citep{Mor2019,Zonoozi2019}.
The description of star-formation with similar conditions as found in the present-day Milky Way is described robustly. Empirical relations describing conditions departing from well-measured regions in the Milky Way neighbourhood are subject to obtaining improved measurements, possibly with future instruments. 
\\
Nevertheless, while the exact shape of the galaxy-wide IMF might change with an improved empirical description of star-formation, the expected trend of the IMF becoming top-heavy/light at high/low SFR
is now supported by a large variety of observations \citep[e.g.][]{Zhang18, Brown2019, Kalari2018, Schneider18, Lee2009, Watts2018, Conroy2017, Martin-Navarro15} 
\\
In this study we use the IGIMF3 model from \citet{Jerabkova18}, 
which implements the variation of the stellar IMF over the entire stellar mass range
(see Fig. \ref{fig:IGIMF_3}).
We refer to the galaxy-wide IMF as described by the IGIMF3 model as the non-universal 
or environment-dependent IMF throughout the rest of this paper.
We briefly discuss the difference in results obtained with the IGIMF2 model 
from \citet{Jerabkova18}, which implements
only variation of the high-mass part
of stellar IMF, in the Appendix \ref{app: IGIMF}. 

\subsection{SFR corrections for the non-universal IMF}\label{sec: method: SFR corr}
\begin{figure}
\centering
\includegraphics[width=\hsize]{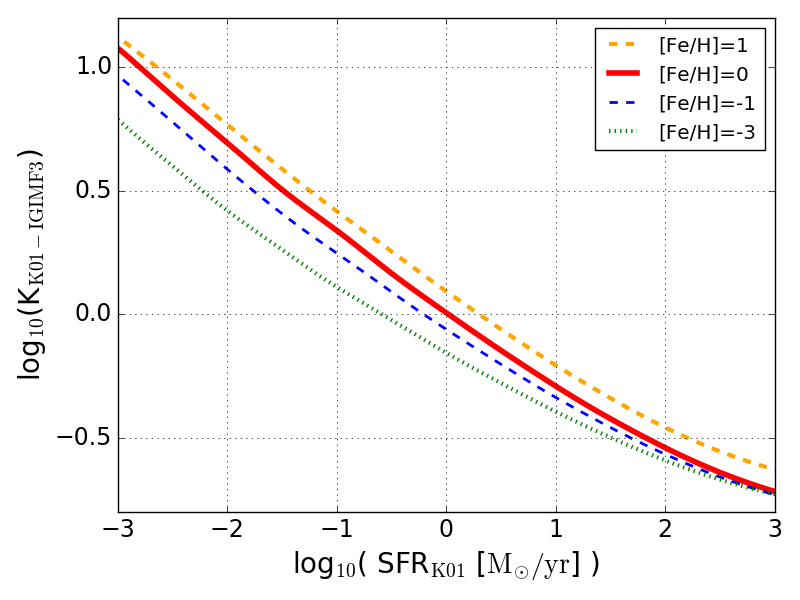}
\caption{
SFR estimated assuming the universal, Kroupa (2001) IMF (SFR$_{\rm K01}$) and the corresponding correction factors (K$_{\rm K01-IGIMF3}$) used to correct that SFR for the environment dependent IMF (described by the IGIMF3 model from \citet{Jerabkova18}). Different lines show the dependence of the correction factor on metallicity (represented by [Fe/H]).
For example, at log$_{10}$(SFR$_{\rm K01}$)=-2 and [Fe/H]=-1, SFR$_{\rm IGIMF3} \approx 10^{0.6}$ SFR$_{\rm K01}$.
}
\label{fig: SFR-corr factor}
\end{figure}

The effect of a non-universal IMF on the $f_{\rm SFR}$(Z,z) 
can be estimated by determining the error one makes in the observation-based galaxy SFR assuming an universal IMF and then correcting for it. 
The observational estimates of the SFR rely on various tracers
that typically measure the number of recently formed massive stars
\citep[e.g. UV luminosity -- either measured directly or in the infrared when reprocessed by dust, H$\alpha$ emission from the recombination of gas ionised by the most massive stars;][]{MadauDickinson14}.
This number can be converted to the total amount of mass turned into stars assuming a particular IMF \citep[e.g.][]{Kennicutt98,KennicuttEvans12,MadauDickinson14}.
\\
For a given SFR tracer, one can calculate the correction factor $K=\frac{SFR_{\rm IGIMF}}{SFR_{\rm K}}$ 
that allows to convert $SFR_{\rm K}$  - the SFR estimate that assumes an universal IMF to $SFR_{\rm IGIMF}$ - the SFR that would 
be measured with a non-universal 
IMF($SFR$, [Fe/H]\footnote{
[Fe/H]=$\rm log_{10}\left(\frac{Fe}{H}\right) - log_{10}\left( \frac{Fe_{\odot}}{H_{\odot}}\right)$, where Fe and H stand for the number density of the corresponding atom.}).
\\
\citetalias{ChruslinskaNelemans19} construct their SFMR by using
the relation found by \citet{Boogaard18} for the local low--to--intermediate mass galaxies
and exploring three variations on the shape of the uncertain high mass end of the relation
(either a single power-law, a broken power-law or a power-law with a sharp flattening, see Fig. 5 in \citetalias{ChruslinskaNelemans19}).
This local relation, which sets the initial normalisation of the SFMR
\footnote{
The normalisation of the SFMR relation in \citetalias{ChruslinskaNelemans19} is assumed to increases with redshift as found by \citet{Speagle14} up to $z$=1.8,
and the rate of its evolution is adjusted to reproduce the observed flattening in the redshift evolution of SFMR normalisation at higher $z$ (see Sec. 2.4 in \citetalias{ChruslinskaNelemans19} and references therein).
We note that the SFMR constructed that way is not directly based on any particular tracer at higher $z$,
nevertheless, we use the matallicity dependent correction of the H$\alpha$--based SFR at all redshifts to obtain a rough
estimate of the effect of the non-universal IMF on the $f_{\rm SFR}$(Z,z).
}
relies on the H$\alpha$ based SFR measurements 
(i.e. the authors employ the \citet{Kennicutt98} relation).
Therefore, we use the matallicity dependent corrections to H$\alpha$--based SFR 
to estimate the effect of the non-universal IMF on the $f_{\rm SFR}$(Z,z).
\\
Such corrections to the \citet{Kennicutt98} relation were recently calculated by  \citet{Jerabkova18}.
Using the PyPegase python wrapper\footnote{ https://github.com/coljac/pypegase \\
We note that the updated version PEGASE.3 has been recently released \citep{Fioc19}. The main difference with respect to PEGASE.2 is the addition of modelling of dust emission and its evolution and corresponding extent of computed wavelengths
and hence it would not affect the results shown in this study.
}, \cite{Jerabkova18} combined the GalIMF code \citep{Yan17, Yan2019code} which computes the galaxy-wide IMF($SFR$, [Fe/H]) with PEGASE.2 stellar population synthesis code \citep[][]{Fioc1999, Fioc2011} to produce the H$\alpha$ flux for a given stellar population assuming a constant star formation history (SFH).
\\
For the purpose of this paper we extended previous calculations for a larger parameter space of SFRs ($10^{-5}\, M_{\odot}/\mathrm{yr}$, $10^3\, M_{\odot}/\mathrm{yr}$) and [Fe/H] metallicities (-5,1). 
\\
The corresponding relation between the universal IMF based SFR
and the SFR correction factor for the underlying non-universal IMF is shown in Fig. \ref{fig: SFR-corr factor} for different [Fe/H] values.

\begin{figure}
    \centering
    \includegraphics[width=\hsize]{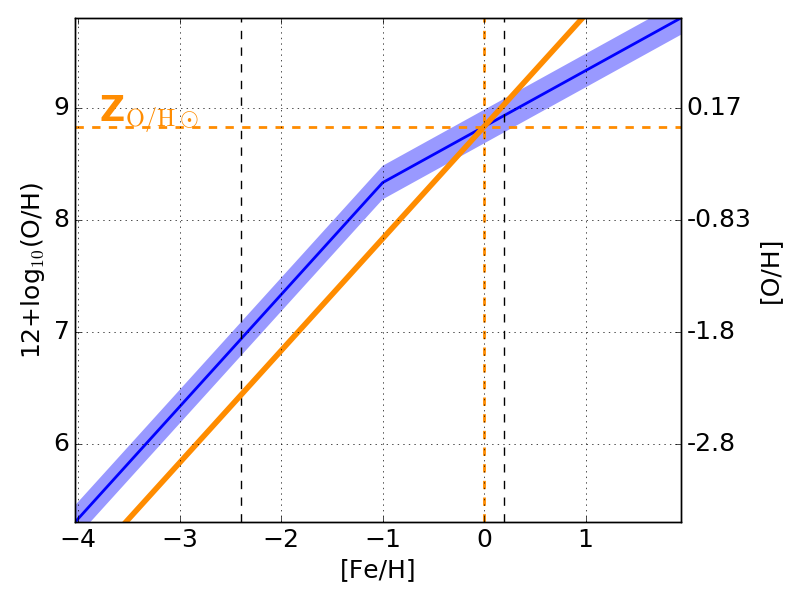}
    \caption{Assumed $Z_{\rm O/H}$ -- [Fe/H] conversion. The blue line results from Eq.~\ref{eq: OFe-FeH} and \ref{eq: FOH-FeH}. We introduce normally distributed scatter ($\sigma$=0.15 dex) around the central relation, indicated by the shaded area (showing 1 $\sigma$ distance from the line). The [Fe/H]=[O/H] relation is shown as the thick orange solid line for the reference. The orange dashed lines indicate the solar metallicity in both measures. The black dashed lines roughly indicate the range of [Fe/H] probed in Milky Way 
    studies (see Appendix \ref{app: FOHFeH}).}
    \label{fig: FOH-FeH_conversion}
\end{figure}

The metallicity dependence of the IMF in \citet{Jerabkova18} is parametrised 
 using the relative iron abundance (relative to solar; [Fe/H]),
 while the observations used in \citetalias{ChruslinskaNelemans19} provide the oxygen abundance $Z_{\rm O/H}$ ($Z_{\rm O/H}=$12+log$_{10}$(O/H)). 
 To combine the two methods, one needs to convert one metallicity measure to the other, which is not a straightforward procedure.
 However, we note that the metallicity dependence of the IMF
has only a secondary effect on the correction factors calculated
by \citet{Jerabkova18} and so the
exact form of the relation does not have a strong effect on
our results.
For the purpose of this study we assume an example relation between [O/H] and [Fe/H]
as shown in Fig. \ref{fig: FOH-FeH_conversion}
that takes into account the overabundance of oxygen with respect to iron 
relative to solar abundances at low metallicities \citep[e.g.][see Appendix \ref{app: FOHFeH} for more details]{WheelerSnedenTruran89,ZhangZhao05,Izotov06,Tolstoy09}.
\\
Throughout this paper we use solar metallicity $Z_{\rm O/H \odot}$=8.83 as reported in \citet{AndersGrevesse89}, falling roughly in the middle of the range of the presently considered values \citep[e.g.][]{DelahayePinsonneault06,Asplund09,Vagnozzi17}.

\subsection{Combining the two methods}
The procedure applied to account for a non-universal IMF can be
summarised as follows:
\begin{itemize}
 \item As in \citetalias{ChruslinskaNelemans19}, 
 we use the GSMF to obtain the number density of galaxies
 of different $M_{\rm gal,\ast}$ ($n_{gal,\ast}$). 
 In general, $M_{\rm gal,\ast}$ estimates are affected by the change
 in the underlying IMF. However, 
 $M_{\rm gal,\ast}$ is only used to connect the various properties of galaxies
 and is not explicitly used in our calculations. 
 Even though the $M_{\rm gal,\ast}$ would be different
 in the non-universal and the universal IMF case, 
 it would correspond to the same number density of objects
 \footnote{
We assume here that the shape of the GSMF would not change 
when we change the IMF. This is true for the universal IMF,
however, may not be strictly true for the non-universal IMF.
The construction of the GSMF involves binning the observed galaxies
with respect to the inferred masses (which depend on the assumed IMF).
One bin contains galaxies that have similar masses, but may differ
in SFR and metallicity. Once the IMF becomes SFR and
 metallicity dependent, the galaxies in principle may 'change bins'.
However, the existence of the scaling relations guarantees that galaxies of
similar masses have similar SFR and metallicity and within the range of
scatter the  non-universal IMF as described by the IGIMF would not affect the masses within one bin in a very different way and so the shape of the
GSMF would not be altered.
 }.
 \item As in \citetalias{ChruslinskaNelemans19}, we assign the SFR to 
 a certain  $n_{gal,\ast}$ (via $M_{\rm gal,\ast}$ and GSMF)
 using the SFMR with scatter.
 The SFR estimate is affected by the change in the underlying IMF.
 We correct for this at later step.
 \item We assign the metallicity $Z_{\rm O/H}$ 
 to a certain  $n_{gal,\ast}$ (via $M_{\rm gal,\ast}$ and GSMF)
 with the MZR, taking into account the scatter and the fundamental metallicity relation as described in \citetalias{ChruslinskaNelemans19}.
 We assume that $Z_{\rm O/H}$ estimates obtained for various metallicity
 calibrations are not significantly affected by the assumed IMF, i.e. if
 the underlying IMF is different than the assumed universal IMF,
 one would measure the same $Z_{\rm O/H}$.
 \item We convert $Z_{\rm O/H}$ to [Fe/H] as discussed in Sec. \ref{sec: method}.
 This step is necessary, as the metallicity dependence of the IMF in \citet{Jerabkova18} is parametrised 
 using the iron abundance as a measure of metallicity.
 \item Knowing the universal IMF-based SFR and [Fe/H] corresponding to a certain
 number density of galaxies,
 we calculate the correction factor for the SFR
 as described in \citet{Jerabkova18} and estimate the corrected SFR.
 \item We repeat the procedure at different redshifts, taking into account
 evolution of the GSMF and the scaling relations, and construct 
 the $f_{\rm SFR}$(Z,z).
\end{itemize}

\section{Results}\label{sec: results}

\subsection{Effect on the $f_{\rm SFR}$(Z,z)}

\begin{figure*}
    \centering
    \includegraphics[width=\hsize]{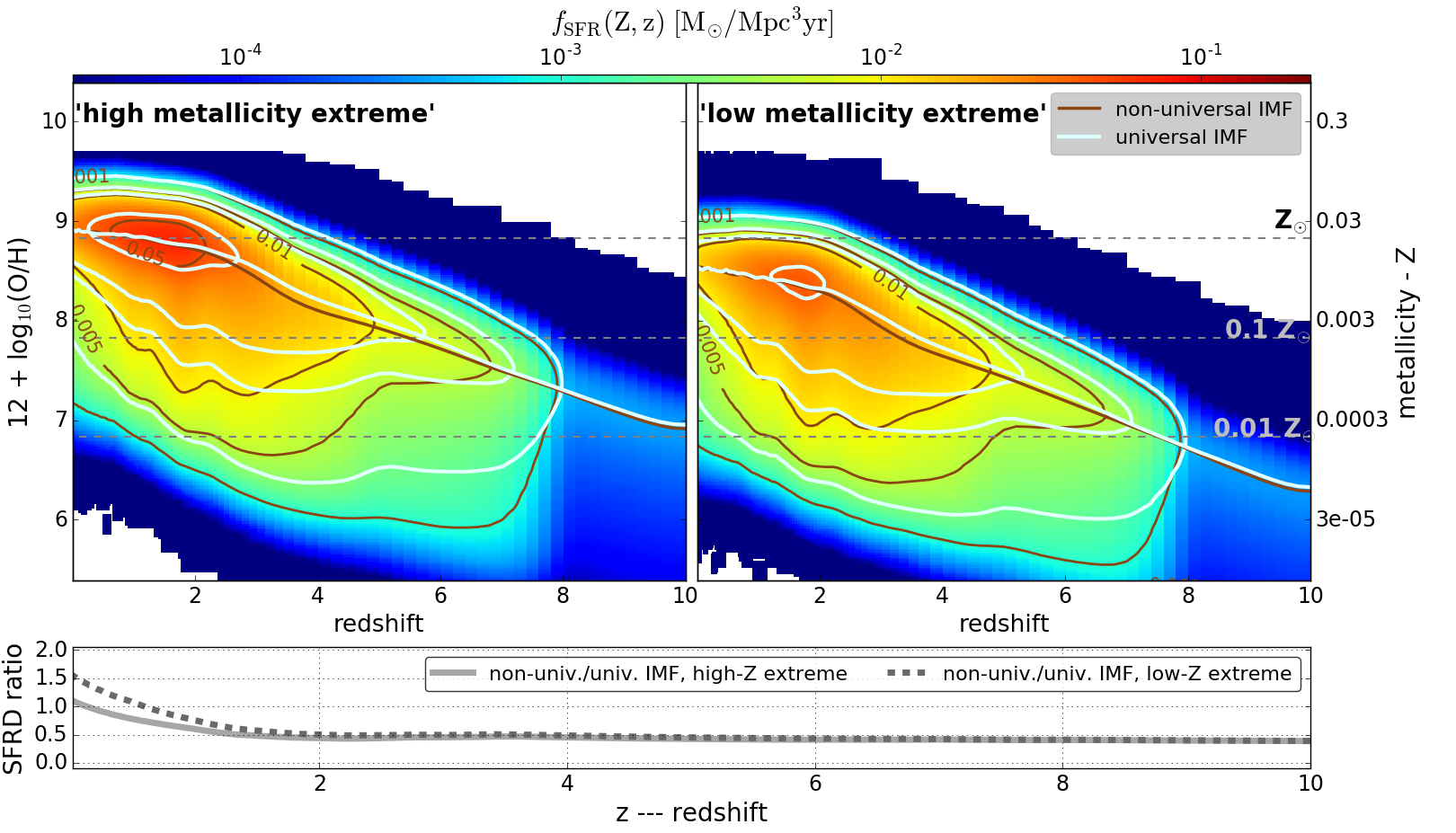}
    \caption{Upper panels: distribution of the SFRD at different metallicities and redshift ($z$). Left --  background colour and brown contours: high metallicity extreme assuming the non-universal IMF; white contours - high metallicity extreme assuming universal IMF. Right -- the same as in the left panel, but for the low metallicity extreme. To plot the white contours, $f_{\rm SFR}$(Z,z) for the universal IMF cases has been re-scaled in such a way that the total SFRD at each redshift matches that of the corresponding variation with the non-universal IMF (see the bottom panel; the non-scaled $f_{\rm SFR}$(Z,z) in the universal IMF cases is shown in Fig. 8 in \citetalias{ChruslinskaNelemans19}). This reveals the presence of the more extended low metallicity tail present in the non-universal IMF variations 
    \newline
    Bottom panel: the ratio of the total SFRD at each redshift as calculated under the assumption of the non-universal to universal IMF for the two $f_{\rm SFR}$(Z,z) variations (dashed line -- low-Z extreme, solid line -- high-Z extreme).}
    \label{fig: SFRD_Z_z_extremes}
\end{figure*}
  
The general effect of the SFR corrections for the environment dependent IMF is to increase the SFR of the low-mass galaxies and decrease that of the massive, highly star-forming galaxies.
This means the galaxy main sequence is less steep and the total SFRD at a given redshift is less dominated by the contributions from the most massive, relatively metal-rich galaxies than what one would expect under the assumption of an universal IMF.
\\
This has two main effects on the $f_{\rm SFR}$(Z,z) distributions,
as can be seen in Fig. \ref{fig: SFRD_Z_z_extremes}.
This figure shows the comparison of the effect of the change
from the universal to non-universal IMF for two
$f_{\rm SFR}$(Z,z) distributions: the high and low metallicity extremes introduced in \citetalias{ChruslinskaNelemans19}.
Those distributions  were chosen as variations of the $f_{\rm SFR}$(Z,z) allowed by the current uncertainties in the empirical scaling relations
that lead to the highest fraction of stellar mass that forms
since $z=3$ at high ($>Z_{\rm O/H\odot}$) and low ($<0.1 Z_{\rm O/H\odot}$) metallicity respectively (see Appendix \ref{app: SFRD_Z_z} for more details).\\
The first effect of changing to the non-universal IMF is the presence of a more pronounced low-metallicity tail in the $f_{\rm SFR}$(Z) distribution at any redshift, which can be seen by comparing the location of the white and brown contours in Fig. \ref{fig: SFRD_Z_z_extremes}.
Those contours show regions of the constant SFRD for several example values in the universal and non-universal IMF case respectively, after eliminating the difference in the total SFRD
(i.e. $f_{\rm SFR}$(Z,z) summed over metallicities at each redshift) between the two cases.
\\
The second effect is on the total SFRD found at
each redshift.
The high-SFR end of the galaxy main sequence is the most strongly
affected by the change in the underlying IMF, which results in a net reduction of the total SFRD at a given redshift when the non-universal
IMF is assumed.
This is demonstrated in the bottom panel of Fig. \ref{fig: SFRD_Z_z_extremes}, which shows the ratio of the total SFRD at different $z$ obtained under the assumption of the non-universal versus
universal IMF.
The ratio becomes more extreme towards higher redshifts, because at relatively high SFRs the magnitude of the applied SFR corrections is larger and
the average galactic SFR increases with $z$.
There is also an additional trend with metallicity, causing the corrections at the high-SFR end to increase with decreasing metallicity (and therefore with increasing $z$, see Fig. \ref{fig: SFR-corr factor}).

\subsection{Effect on the formation of stars of different masses}

\begin{figure}
    \centering
    \includegraphics[width=\hsize]{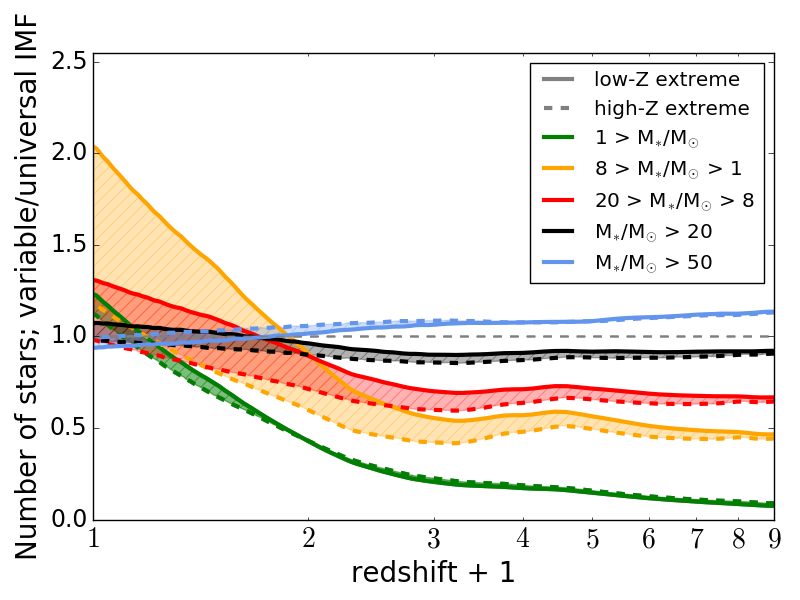}
    \caption{Ratio of the number of stars forming in different mass ranges in the case with the environment dependent IMF to the universal IMF as a function of redshift. The shaded area spans between the ratios obtained for the low (solid lines) and high (dashed lines) metallicity extreme $f_{\rm SFR}$(Z,z) distributions from \citetalias{ChruslinskaNelemans19}.}
    \label{fig: N_ratios_IGIMF3}
\end{figure}
  
\begin{figure}
    \centering
    \includegraphics[width=\hsize]{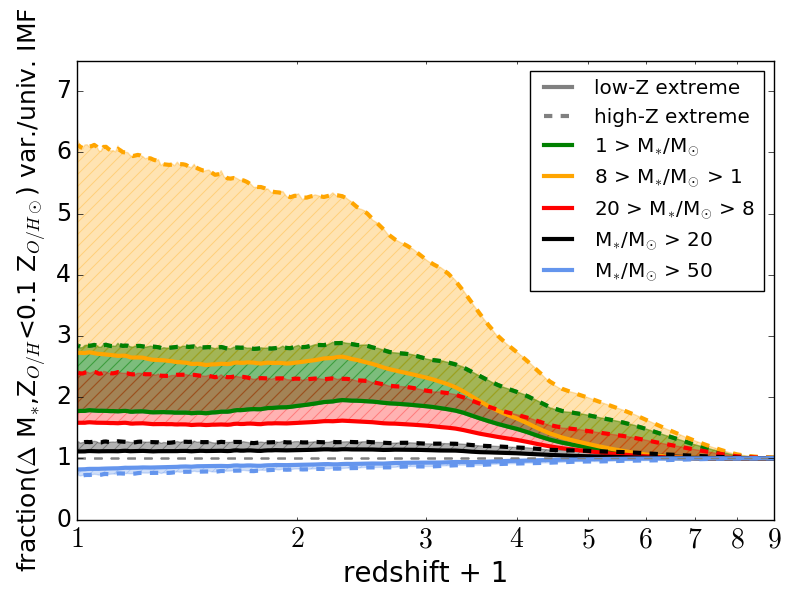}
    \caption{Ratio of the fraction of stars forming at low metallicity ($Z_{\rm O/H}<Z_{\rm O/H\odot}$) in different mass ranges in the case with the environment dependent IMF to the universal IMF as a function of redshift. The shaded area spans between the ratios obtained for the low (solid lines) and high (dashed lines) metallicity extreme $f_{\rm SFR}$(Z,z) distributions from \citetalias{ChruslinskaNelemans19}.
    Thus, at redshift $z$=1, about 2.5 times more stars with M$_{*}<1 \Msun$ form in the considered non-universal IMF scenario than in the universal IMF case.
    }
    \label{fig: NlowZ_ratios_IGIMF3}
\end{figure}

\begin{figure*}
    \centering
    \includegraphics[width=\hsize]{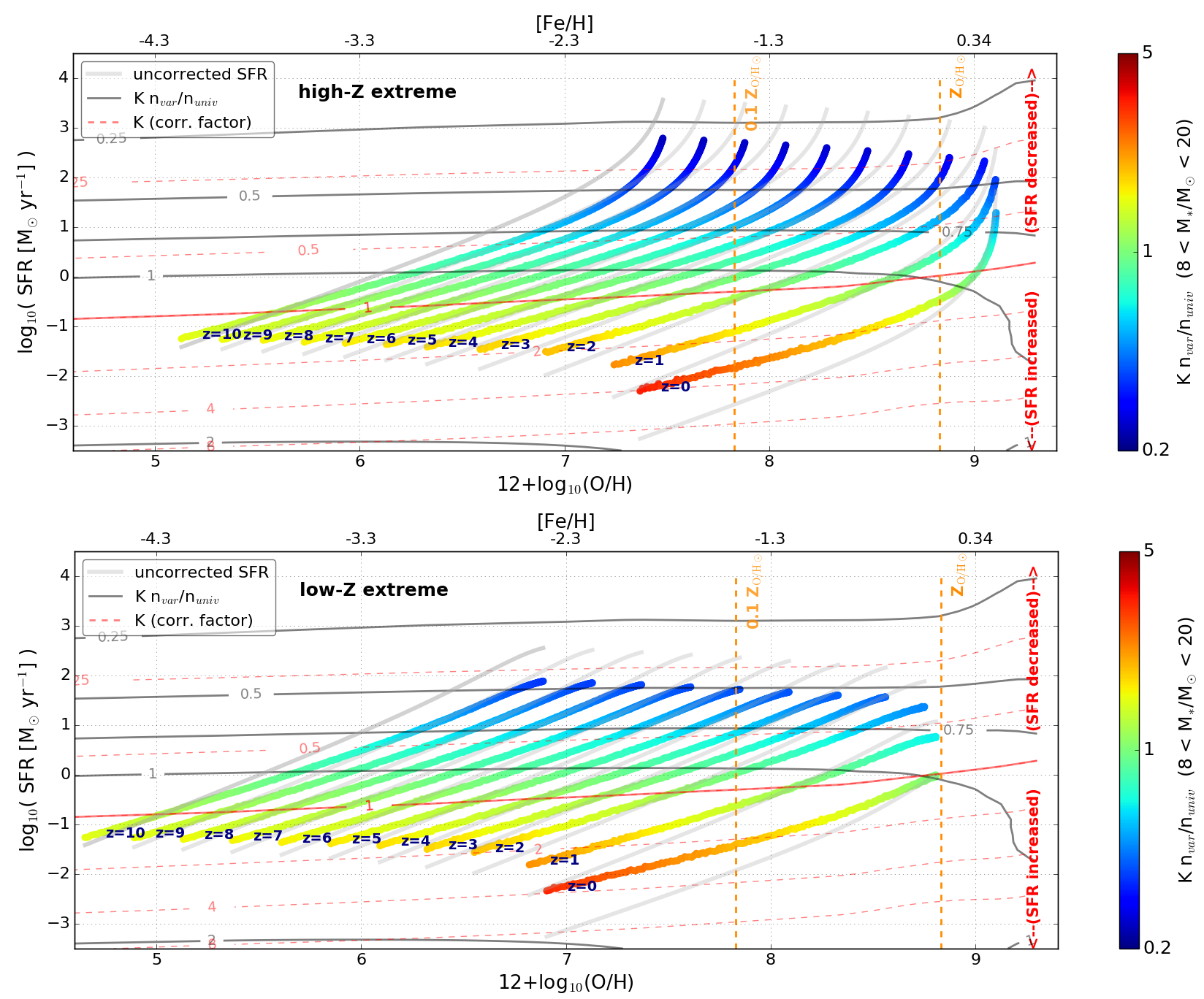}
    \caption{Effect of the transition from the universal to environment dependent IMF on the formation of the NS progenitors. The grey thick curves show the SFR--metallicity ($Z_{\rm O/H}$) relation as used in the high-Z extreme (top) and low-Z extreme (bottom) $f_{\rm SFR}$(Z,z) from \citetalias{ChruslinskaNelemans19} at different redshifts. The top metallicity axis ([Fe/H]) was obtained from $Z_{\rm O/H}$ using the conversion described in Sec. \ref{sec: method}. The coloured curves show the same relation after correcting the SFR for the non-universal IMF (see Fig. \ref{fig: SFR-corr factor}). The colour denotes the ratio of the number of stars forming in the NS progenitor mass range in the non-universal vs universal IMF case. The near-horizontal grey lines are the lines of constant value of that ratio. The red dashed lines are the lines of the constant correction factor. The SFR below the thick red solid line is increased and above that line is decreased with respect to the universal IMF case. The vertical orange dashed lines show solar and 10\% solar metallicity.}
    \label{fig: IGIMF3}
\end{figure*}

The assumption about the IMF
also affects the number of stars of different masses ever formed, as well as the 
distribution of metallicities at which those stars have formed.
\\
The ratio of the \textit{number of stars} formed in different mass ranges in the variable versus
universal IMF case is shown in Fig. \ref{fig: N_ratios_IGIMF3}.
Fig. \ref{fig: NlowZ_ratios_IGIMF3} demonstrates the corresponding ratio of the \textit{fraction of stars} forming at low metallicity (below 10\% $Z_{\rm O/H \odot}$) in different mass ranges.
The ratios are shown for a few mass ranges that correspond to the 
conventionally adopted white dwarf (WD; $M_{*}<8 M_{\odot}$), neutron stars (NS; $8 M_{\odot} < M_{*}<20 M_{\odot}$) and black hole (BH; $M_{*}>20 M_{\odot}$) progenitor mass limits
\footnote{
We note that those ranges are used for the convenience of discussion and for the illustration purposes.
The exact mass limits of the progenitors of different types
of stellar remnants are uncertain and depend e.g. on metallicity, rotation,
presence of binary interactions \citep[e.g.][]{Heger03,Ekstrom12,Ibeling13,Doherty15}. Especially the boundary between BH
and NS progenitors can differ significantly from the assumed limit and the recent simulation results suggest there may well not be a set mass boundary between them \citep[e.g.][]{SukhboldWoosley14,Sukhbold16}.
}.
We additionally separate the WD progenitors mass range below and above 1 $M_{\odot}$\footnote{There are a few reasons for separating these mass ranges. 1 $M_{\odot}$ is the conventional mass limit at which one of the IMF power law breaks occurs, separating the 'high mass' and the 'low mass' part of the IMF. Also, the difference between the IGIMF2 or IGIMF3 models from \citet{Jerabkova18} lies in the description of the low mass $<1 M_{\odot}$ part of the IMF (see Appendix \ref{app: IGIMF}). Furthermore, most of the stars formed below that mass are still 'to-be WDs', with the evolutionary timescale approaching the age of the Universe. 
}
and plot the ratios for the progenitors of the most massive stellar origin BHs from isolated stellar/binary evolution,
i.e. stars of initial masses $M_{*}>50 M_{\odot}$.\footnote{
A $\approx$50 $\rm M_{\odot}$ star evolving in isolation at $\approx$10\% solar metallicity is expected to leave behind a $\approx$20-40 $M_{\odot}$ BH \citep[e.g.][]{Heger2003,SperaMapelli17,MandelFarmer18}.
}
In general, the magnitude of the effect of the change in the IMF
on the formation of stars in different mass ranges also depends 
on $f_{\rm SFR}$(Z,z). 
To demonstrate this, in Fig. \ref{fig: N_ratios_IGIMF3} and
Fig. \ref{fig: NlowZ_ratios_IGIMF3} we show the ranges spanning between the
ratios obtained using the limiting cases from \citetalias{ChruslinskaNelemans19},
i.e. the low (solid lines) and high (dashed lines) metallicity extreme $f_{\rm SFR}$(Z,z) distributions. 
\\
Before we zoom into the individual mass ranges, we discuss some general trends
that appear in in Fig.~\ref{fig: N_ratios_IGIMF3} and \ref{fig: NlowZ_ratios_IGIMF3}:
\begin{itemize}
    \item At high $z$ the overall number of stars forming in the non-universal IMF scenario is smaller than in the universal IMF case.
    \item The ratios in Fig. \ref{fig: N_ratios_IGIMF3} increase towards lower $z$, and exceed unity at $z<1$ if the $f_{\rm SFR}$(Z,z) distribution is similar to the low-Z extreme from \citetalias{ChruslinskaNelemans19}.
    \item The fraction of stars forming at low metallicity is increased in the non-universal IMF scenario. This fraction increases towards low $z$.
\end{itemize}
The above statements are common to all but the highest mass range ($M_{*}>50 M_{\odot}$) considered, which shows the reverse behaviour. We discuss that in Sec.~\ref{sec: BH prog}. The trends seen in the remaining mass ranges can be understood by looking at 
Fig.~\ref{fig: IGIMF3}. The thick-light-gray curves in this figure show the SFR--metallicity relation at different redshifts as used in the high-Z extreme (top panel) and low-Z extreme (bottom panel) $f_{\rm SFR}$(Z,z) from \citetalias{ChruslinskaNelemans19}.
The colored solid curves show that same relation after correcting the SFR for the non-universal IMF as discussed in Sec. \ref{sec: method: SFR corr}. The colour shows the \textit{ratio} of the number of stars formed in the NS progenitor mass range in the non-universal to universal IMF scenario in the different regions of the SFR--metallicity plane.
\\
It can be seen that at high redshifts the coloured curves fall nearly entirely below the grey curves - i.e. the SFR of almost all galaxies is lowered with respect to the universal IMF case. This results in a decreased number of stars forming in the 
non-universal IMF scenario (but the strength of this effect differs for different masses of stars as shown in Fig. \ref{fig: N_ratios_IGIMF3}).
\\
As the redshift decreases, the average SFR--metallicity relation shifts towards higher
metallicities and lower SFR. This leads to higher positive SFR corrections at the low SFR side of the relation (see Fig. \ref{fig: SFR-corr factor}) and the increasing fraction of the low-SFR part of the coloured curves appears above the corresponding grey curves. At the same time, the decrease on the high SFR end of the relation due to the correction is smaller. Hence, the ratio of the number of stars forming in the non-universal to the universal IMF scenario increases towards lower redshifts.
\\
For the low-Z extreme $f_{\rm SFR}$(Z,z), nearly the entire corrected relation at low $z$
falls above the corresponding universal-IMF relation - i.e. the SFR of almost all galaxies is higher in the non-universal IMF scenario. In that case the number of stars forming in different mass ranges exceeds that expected under the assumption of the universal IMF. In the high-Z extreme the galaxies with the highest universal IMF SFRs receive negative corrections even at $z\approx0$ and hence the increase in the fraction of the number of stars forming in the non-universal to universal IMF scenario is smaller when this  $f_{\rm SFR}$(Z,z) distribution is used. 
\\
The increase in the \textit{fraction} of stars forming at low metallicity in the non-universal with respect to the universal IMF scenario is also readily seen in Fig. \ref{fig: IGIMF3}. The corrected SFR is always decreased on the high metallicity and increased on the low metallicity side of the SFR--metallicity relation. The fractions shown in Fig. \ref{fig: NlowZ_ratios_IGIMF3} correspond to metallicity below 10\% solar and so at $z>$8 (6) in the high (low) metallicity extreme the ratio of the fractions
of stars forming below that threshold is close to unity (nearly all star formation happens below 10\% solar metallicity i.e. to the left of the left-most orange vertical line in the Fig. \ref{fig: IGIMF3}). The ratio then increases towards lower redshifts,
as the SFR shifts to higher metallicities and is decreased above and increased below $0.1 Z_{\rm O/H}$ in the non universal IMF scenario. The decrease at high $Z_{\rm O/H}$ is stronger at higher $z$, while the increase at low $Z_{\rm O/H}$ is stronger at lower $z$
and the increase in the ratio starts to flatten between $z$=1 and 2, where the two effects balance each other. The ratio is smaller if the low-Z extreme $f_{\rm SFR}$(Z,z) is used, as in this case even at low redshifts a significant fraction of the total SFRD falls below 10\% solar metallicity, irrespective of the assumptions about the IMF discussed in this study.

\subsubsection{BH progenitors}\label{sec: BH prog}

The effect on the number and fraction of the most massive stars 
(BH progenitors) forming at low metallicity is shown by the black 
(stars with the initial masses $M_{*}>$20 $M_{\odot}$) 
and blue ($M_{*}>50 M_{\odot}$) lines in Fig. \ref{fig: N_ratios_IGIMF3} and
Fig. \ref{fig: NlowZ_ratios_IGIMF3} respectively.
In general, this effect is very small. This should be expected, as those stars, if present within a galaxy, are responsible for the emission of the bulk of the radiation that serves as an observable SFR tracer.
In particular, the contribution to the H$\alpha$ emission (considered as SFR
tracer in this study) comes only from the most massive stars (O and early-type B-stars with $M_{*}\gtrsim 17 M_{\odot}$, e.g. \citealt{MadauDickinson14}).
This means that stellar population forming according to a top-heavy IMF would have higher H$\alpha$ luminosity per unit mass of star formed than the one following the canonical MW-like IMF. Both IMFs can be used to interpret the measured luminosity, as long as the inferred SFR is adjusted accordingly (i.e. it needs to be lower in the top-heavy IMF case). Therefore, the effect of the high mass end slope of the IMF and the SFR counterbalance each other and the total number of the most massive stars formed, as well as their metallicity distribution is not strongly affected by the IMF variations.
\\
While the number of all stars more massive than 20 $M_{\odot}$ forming at z$\gtrsim$1
is slightly decreased, the number of those with the highest masses is higher in the
non-universal IMF case.
This is because the IMF is more top-heavy in the high-SFR and low-metallicity environments - typical of high redshift galaxies, and the contribution to H$\alpha$ luminosity increases with stellar mass.
We find that the formation of the massive ($\gtrsim 30 M_{\odot}$) 
stellar origin BH from isolated stellar evolution in the low-redshift
Universe is disfavoured. The progenitors of those BHs are believed to form at low metallicity with a mass $\gtrsim 50 M_{\odot}$.
This is required to avoid strong stellar wind mass loss expected at high metallicities \citep[e.g.][]{Vink01,VinkDeKoter05} and retain enough mass during their evolution to form a massive remnant (e.g. \citealt[][]{Heger03,Belczynski10Mmax,SperaMapelli17}).
At low redshifts the star formation with such low metallicities happens predominantly in dwarf galaxies, forming stars at low rates $(\lesssim 0.1 M_{\odot} /\rm yr)$. 
Those galaxies are deficient in massive stars as indicated by their H$\alpha$/UV emission \citep{Pflamm2009,Lee2009,Meurer2009} and the photometry of resolved stellar populations \citep{Watts2018}. This is consistent with the prediction of the adopted IGIMF theory, which leads to the galaxy-wide IMF in the low SFR (low mass) galaxies that is top-light and truncated at lower $M_{*}$ when compared to the universal IMF (see e.g. Fig. \ref{fig:IGIMF_3}). Thus, those galaxies are less probable hosts of the massive stellar origin BHs.

\subsubsection{NS progenitors}
The results for the stars forming with masses in the conventional
NS progenitor mass range of 8-20 $M_{\odot}$ follow the trends discussed
earlier in this section and are indicated in red in
Fig. \ref{fig: N_ratios_IGIMF3} and Fig. \ref{fig: NlowZ_ratios_IGIMF3}.
The number of stars forming in this mass range is generally decreased, except for the low redshift range $z\lesssim 1$ when the $f_{\rm SFR}$(Z,z) is described with the
distribution similar to the low-Z extreme from
\citetalias{ChruslinskaNelemans19}.
This decrease reaches a factor of $\approx$1.3 at high $z$
and is stronger than in the BH progenitors mass range, as the
decrease in the SFR of the active galaxies due to the necessary corrections for the non-universal IMF is not fully counterbalanced by the change in the IMF in this mass range.
The fraction of NS progenitors forming at low
metallicity is increased by a factor of $\approx$2 at low to intermediate redshifts.

\subsubsection{WD progenitors}\label{sec: WD progenitors}

The change from the universal to environment dependent IMF has the strongest
effect on the number and metallicity of the WD-progenitors.
This is shown as orange (for initial masses 1 to 8 $M_{\odot}$) and green
(for $M_{*}<1 M_{\odot}$) areas in Fig. \ref{fig: N_ratios_IGIMF3} and 
\ref{fig: NlowZ_ratios_IGIMF3}.
The fraction of stars in the former mass range forming at low metallicity is
up to a factor of $\approx$3-6 higher (depending on the $f_{\rm SFR}$(Z,z)) at $z\lesssim$1.5 in the non-universal than in the universal IMF scenario.
The overall number of the WD progenitors with $M_{*}>1 M_{\odot}$ is 
decreased with respect to the universal IMF scenario by a factor of 
$\approx$2 at $z\gtrsim$1.5 and increased at low redshifts
by up to a factor of 2 if the low-Z extreme $f_{\rm SFR}$(Z,z) is used.
We discuss the importance of those effects for the formation
of type Ia supernovae (SN) progenitors in Sec. \ref{sec: discussion: type Ia}
\\
The number of the lowest mass stars $<1 M_{\odot}$ is decreased by more than a factor of 10
at high $z$ in the considered model of the IMF variations with respect to the universal IMF case.
This is because the formation of low-mass stars is expected to be suppressed in the metal-poor
and hot early Universe. 
This effect is expressed in the IGIMF formulation with a strong metallicity dependence of
the low mass part of the IMF, which is such that the IMF becomes increasingly bottom-light
with decreasing metallicity (more relevant for the star formation at high redshifts)
irrespective of the SFR (and vice versa in the environments with metallicities exceeding
the solar value). 
The fraction of stars with $M_{*}<1 M_{\odot}$ forming at low metallicity is still increased with respect to that expected in the universal IMF scenario due to smaller reduction (at high $z$) or higher increase (at low $z$) in the SFR of the low-metallicity galaxies than in the SFR of the more metal-rich galaxies due to the applied corrections (see Fig. \ref{fig: SFR-corr factor}).
We note that those two effects (i.e. decrease in the number of low-mass stars forming at high $z$ and increase in the fraction of those forming at low metallicity at lower $z$) have the opposite effect on the metallicity distribution of all the low-mass stars ever formed. 
Therefore, this distribution is not significantly different between the considered universal and non-universal IMF scenarios. However, the age distribution of the low-mass, low-metallicity stars ever formed that can still be observed in the local Universe would differ between the universal and non-universal IMF scenarios. In the latter case the bulk of the low-mass, low-metallicity stars forms at later times in the cosmic history and therefore their age distribution would be shifted to younger ages.
\\
We note that the observational constraints on the metallicity dependence of the low mass ($<1 M_{\odot}$) end slope of the IMF are currently very limited \citep{Kroupa02, Marks2012TH}.
If this dependence is much weaker than assumed in the adopted IMF model, the results shown for the stars in this mass range would be closer to those shown for the more massive WD progenitors.
This is shown in the Appendix \ref{app: IGIMF}, where we additionally discuss the difference in the results that would be obtained with no metallicity dependence of the low mass part of the IMF (i.e. the IGIMF2 model from \citet{Jerabkova18}). 

\section{Discussion}

The presented results assume a particular parameterisation of the IMF
dependence on SFR and metallicity.
This description of the IMF variations has been validated
mainly on the sample of relatively low redshift galaxies \citep[e.g.][]{Yan2019}.
We note that the exact form of this dependence is likely to improve
with improving empirical constraints, especially at higher redshifts, 
but the expected \emph{trends} 
in the shape of the IMF (e.g. top/bottom-heaviness at high/low SFR)
with the environment are now reported in many observational studies 
and thus can be expected to stay.
\\
Therefore, our study should be treated as a qualitative discussion
of the expected impact of the assumptions about the (non)universality
of the IMF on the $f_{\rm SFR}$(Z,z) and the formation of stars in various mass ranges
rather than a robust quantitative estimation of this effect.
It can provide a guidance on whether the calculations of e.g. 
the rates of various stellar evolution related phenomena 
(such as various types of SN) performed
under the assumption of the universal IMF are likely to under/over- predict
the estimated quantity and on the order of magnitude of this effect.
\\
Besides the exact description of the IMF variations, our results
are prone to uncertainties related to the calculations of the
SFR correction factors, i.e. 
conversion between the different metallicity measures, 
the stellar tracks used to perform the 
stellar population synthesis (in particular the uncertainties related to the
treatment of the stellar winds, rotation and binarity) 
and the adopted SFH of galaxies.

\subsection{Metallicity conversion}\label{sec: discussion: metallicity}

In order to combine the IGIMF model with the $f_{SFR}$(Z,z) from \citetalias{ChruslinskaNelemans19} and calculate the SFR correction factors,
we need to convert $Z_{O/H}$ to the [Fe/H] metallicity measure. 
We achieve this by adopting a particular relation, 
as shown in Fig. \ref{fig: FOH-FeH_conversion}.
However, there is no universal way to translate the oxygen to iron abundance.
\\
To demonstrate the impact of this assumption on our results,
we additionally consider the simplest conversion [Fe/H]=[O/H] 
(represented by the thick orange line in Fig. \ref{fig: FOH-FeH_conversion},
i.e. the iron abundance translates directly to oxygen abundance maintaining the
solar ratios).
\\
In that case, the estimated SFR of the low-metallicity, 
low SFR galaxies would be higher than in the case when the conversion from Sec. \ref{sec: method} is applied.
This is because the necessary SFR corrections for the variations in the underlying IMF are smaller at low metallicities and the effect of metallicity on the correction factor is more pronounced at low SFRs.
Hence the fraction of stars forming at low metallicity is increased when the [Fe/H] = [O/H] conversion is used (see Fig. \ref{fig: NlowZ_ratios_comparison}).
The overall effect is relatively small compared to that of the
assumed $f_{SFR}$(Z,z).
The effect of the assumed metallicity conversion 
on the ratio of the number of stars forming in different mass ranges in the universal 
to non-universal IMF scenario is negligible (see Fig. \ref{fig: N_ratios_comparison}).
\\ 
The realistic conversion would likely show some form of $\alpha$-enhancement
at low metallicities and high SFR and in this respect resemble more the one
introduced in Sec. \ref{sec: method} 
rather than the simple [Fe/H]=[O/H] conversion (see Appendix \ref{app: FOHFeH}). 
We therefore conclude that this assumption does not have a significant
effect on our results.

\subsection{Stellar population synthesis}

In order to estimate the H$\alpha$ luminosity produced
by a certain stellar population,
one needs to model the stars that make up the
population together with the 
light and spectra coming from their atmospheres.
This is achieved with the use of stellar population synthesis (SPS) models
\citep[e.g.][]{Leitherer1999,Fioc1999,BruzualChalot03,Fioc2011,Fioc19}.
Different models differ in the implemented stellar/atmospheric models,
but generally lead to comparable results \citep[e.g.][]{Jerabkova17}.
\\
However, a number of effects (rotation, binary evolution) that
are relevant for the modelling of the UV part of the spectrum  
are not taken into account in the commonly 
used SPS models, including the one used in our analysis.
These effects are also not taken into account in the observational studies estimating the galactic SFR used in \citetalias{ChruslinskaNelemans19}.
In general, modelling  binary  systems  has  similar  effects on the spectra
 to including rotation in single star models  \citep{Eldridge2009}.
Rotating stellar models show increased lifetimes, 
effective temperatures and mass loss rates
when compared with the non-rotating ones \citep{Ekstrom12}.
In the case of binaries, interactions can e.g. remove the outer layers of stellar envelopes, rejuvenate the accreting star
or lead to the merger of two stars, all effects leading to increased emission in the blue part of the spectrum with respect to single models \citep[e.g][]{Eldridge2009,Eldridge2017,Gotberg19}.
Both the effects of rotation and binarity increase the number of Wolf–Rayet stars and decrease the number of red supergiants present in the population
(via increased mass loss rates due to rotation or envelope stripping during mass transfer in binaries) and can increase  the  number  of  massive  main-sequence stars observed (rotation via extending the lifetime by mixing more fuel into the core, binaries due to mass accretion) \citep[e.g.][]{Eldridge2009}.
\\
Predicted spectral energy distributions tend to be "bluer" 
when the effect of binaries or rotation is considered
and the corresponding stellar populations produce more UV/H$\alpha$ flux for
the same SFR with respect to those composed of single, non rotating stars
(leading to overestimated SFR based on those tracers).
\citet{Gotberg19} find that the UV luminosity is not significantly 
affected by the presence of stars stripped off their envelopes in binary interactions in stellar populations in which star formation is ongoing or that are younger than about a few 100 Myr, but it may still be affected by the presence of other products of binary interaction.
We note that the most rapid rotation is likely achieved in binaries 
\citep{deMink13} and those effects have not been considered simultaneously.
The more in-depth discussion of their potential effect on our results presents a challenge in itself and is beyond the scope of this study.

\subsection{SFH assumption}

In order to compute SFR--H$\alpha$ relations with the variable galaxy-wide IMF 
(and the SFR correction factors) we assume that the SFH is constant on 
> 5 Myr timescales \citep{Jerabkova18}. 
This assumption ensures constant production of H$\alpha$ by the stellar population. 
The H$\alpha$ flux, being driven by the production of ionising photons,
strongly depends on the population of massive stars ($M_{\ast} > 17 \ M_{\odot}$; we note that even though a 17 $M_\odot$ star has a lifetime of a few 10 Myr, the galactic H$\alpha$ luminosity is dominated by the most massive stars with the shortest lifetimes) and thus consequently also on the SFH. For example in case of a single burst event, H$\alpha$ flux values can differ by few orders of magnitude in a few Myr when the most massive stars evolve. 
The assumption of constant SFH on several Myr time-scale is thus needed in order to provide any SFR--H$\alpha$ relation
\citep[e.g.][]{Kennicutt98,Pflamm-Altenburg07,Pflamm2009}.\\
The assumed SFH can be particularly 
important for the SFR estimates in galaxies with
small values of SFR relative to the Milky Way.
Those galaxies form only a very small number of high-mass (H$\alpha$ producing), short-lived stars. 
Clearly, the estimated SFR will be different
depending on whether the star is caught alive or after it
completed its evolution.
The proper evaluation of this effect it is up
to the future studies.
See \citet{Pflamm-Altenburg07} and \citet{Jerabkova18}
for further discussion of this problem.

\subsection{Type Ia supernovae progenitors and supernovae rates}\label{sec: discussion: type Ia}

\begin{figure}
    \centering
    \includegraphics[width=\hsize]{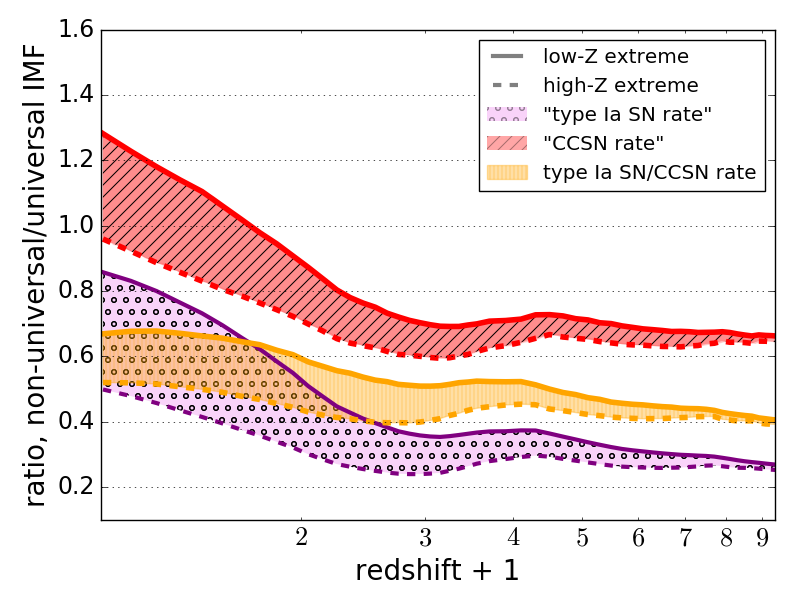}
    \caption{
Comparison of the results of the example calculation of the
CCSN, type Ia SN and type Ia SN to CCSN rates
in the non-universal to universal IMF scenario.
See text for the detailed assumptions.
The colored areas span between the ratios obtained for calculations performed using the low and high metallicity extreme $f_{\rm SFR}$(Z,z) distributions.
}
    \label{fig: Ia_CCSN_ratio}
\end{figure}

The biggest impact of the transition from the universal to the environment dependent IMF is on the formation of stars in the WD progenitor mass range \citep[e.g.][]{WeidnerKroupa05}.
The subset of those stars will be progenitors of the type Ia SN, which are believed to be thermonuclear explosions of relatively massive carbon-oxygen WD. 
There are two main proposed type Ia SNe formation channels, that have evolved over time. 
The first is the accretion model (evolved from the "single degenerate channel")
 in which the explosion is triggered by accretion onto a WD, either by increasing the central density to ignition, 
or by triggering an explosion in the accreting layer that subsequently causes the accreting WD to explode. 
The second is the merger mode (evolved from the "double degenerate channel") in which the explosion is triggered by the merger of two WDs. 
The explosion can be the result of the production of an unstable object that ignites due to a high temperature or density in some parts, 
or as a result of the merging itself and the shocks created in the process. See e.g. \citet{Maoz14}, \citet{LivioMazzali18} 
for a detailed description of these models.
Regardless of the exact formation scenario, the number of 
type Ia SN events expected from a certain 
stellar population depends on the IMF 
\citep[see e.g. Sec. 4.5 in][]{Yan2019}
and will be different in the universal
and in the non-universal IMF scenario.
The assumptions about the IMF also affect the relative rates of different transients of stellar origin
- we discuss type Ia SN and core collapse SN (CCSN) as an example below.
\\
For the illustrative purposes, here we follow the approach described in \citet{Yan2019} and assume that the rate of type Ia SN 
is proportional to the number of stars forming in the mass range suitable for the formation of the primary star in the type
 Ia SN progenitor system times the probability that a companion star also forms in the correct mass range. 
Both quantities are simply calculated from the IMF. 
We assume that the efficiency of formation of type Ia SN from 
progenitor binaries is constant and independent of the assumptions about the IMF.
We are thus neglecting all the effects coming from the possible correlations
between the initial binary parameters, binary evolution and metallicity.
The dominant contribution to type Ia SN in the commonly
considered formation channels is found to come from progenitor stars with the masses in the range 
$\approx 3 - 8 M_{\odot}$ \citep[e.g.][]{Clays14}.
Therefore, we assume that both stars in the type Ia progenitor binary come from this mass range.
The analogous results to those shown in Fig. \ref{fig: N_ratios_IGIMF3} and 
\ref{fig: NlowZ_ratios_IGIMF3}, but for 3-8 $\Msun$ mass range of WD progenitors are shown in the Appendix \ref{app: SNIa}.
Furthermore, we assume that the delay time distribution follows a single power law $\propto t^{-1}$, with a minimum time of 40 Myr
as suggested by observations \citep{MaozMannucci12}.
\\
For CCSNe we simply assume that the rate is proportional to the number of stars forming in the mass range between 8 and 20 $\Msun$,
again neglecting any metallicity or binary evolution-related effects (hence the ratio of the CCSN rate in the non-universal to universal IMF scenario is
identical to the red range in Fig. \ref{fig: N_ratios_IGIMF3}).
\\
The ratio of rates calculated under those
assumptions in the non-universal to universal IMF scenario
is shown in Fig. \ref{fig: Ia_CCSN_ratio}. 
SN rates to first approximation follow the SFR, 
and so can be expected to increase less steeply with $z$ 
towards the peak of the SFH of the Universe 
in the non-universal than in the universal IMF scenario 
(see e.g. Fig. \ref{fig: SFRD_comparison_appendix} 
for the cosmic SFH obtained for different IMF and $f_{\rm SFR}$(Z,z)
assumptions used in this study).
Except for the offset due to longer delay time for type Ia SN explosions,
their rate shows similar behaviour with redshift to CCSN and so there is relatively little evolution with redshift in the type Ia SN to CCSN rate ratio.
The relative rate of type Ia SN to CCSN, as well as type Ia SN rate
as  shown in Fig. \ref{fig: Ia_CCSN_ratio} is lowered in the non-universal IMF scenario by a factor of a few, but we stress that the exact magnitude of this effect depends e.g. on the assumed delay time distribution of type Ia SN.
We note that the formation efficiency for type Ia SN is not known theoretically but inferred from the observations. Therefore, in order to explain the observed type Ia SN rate with a given progenitor model, this efficiency would simply need to be higher in the non-universal IMF than in the universal IMF scenario.
We note that at high $z\gtrsim 3$ the presented trends are subject to additional
uncertainty due to the uncertain contribution of dwarf galaxies to the cosmic SFRD (see Appendix \ref{app: discrepancy}) .
\\
The above considerations neglect any potential metallicity effects on type Ia SN rates.
However, the fraction of stars forming at low metallicity
in the assumed type Ia SN progenitor mass range is increased by a factor of 
$\approx 2 - 4$ (depending on the $f_{\rm SFR}$(Z,z) and IGIMF model; see  Fig. \ref{fig: ratio_type_Ia}) at $z\lesssim 2$ in the non-universal IMF scenario.
This increase may be particularly important if the
efficiency of formation of type Ia SN
depends on metallicity.
In fact, some metallicity dependence has been
reported in theoretical studies
e.g. \citet{Toonen12} find that the formation efficiency
of type Ia SN in the double degenerate channel 
can be up to 30\%-60\% higher at low metallicity 
than at solar-like metallicities
(the authors compare the models at solar and 5\% solar metallicity).
Potential correlation between the SN Ia rate
and galaxy metallicity has also been reported \citep[e.g.][find enhanced SNIa rate in star forming galaxies with lower metallicities]{Cooper09}.
Such correlation might be explained by the metallicity dependence of
the IMF expected within the IGIMF theory.  

\section{Conclusions}

In this study we evaluate the effect of the possible IMF variation on the distribution of
the SFRD over metallicities and redshift.
We also discuss the impact of those variations on the expected number and metallicity of WD, NS and BH progenitors forming throughout the cosmic history.
\\
To this end, we use the empirically driven description of the SFR and metallicity
dependent IMF based on the IGIMF-theory
together with the SFR corrections from \citet{Jerabkova18}.
We apply them to correct the observation-based $f_{\rm SFR}$(Z,z) from \citetalias{ChruslinskaNelemans19} for the underlying environment dependent IMF. 
\\
While the recent observational and theoretical studies indicate clear trends in the 
changes of the shape of the IMF with SFR and metallicity,
the exact dependence of the IMF on the environment may yet find some revision.
We therefore focus mainly on the qualitative effect of the applied corrections.
Our  main  conclusions  are summarised below:
\begin{enumerate}
    \item  $f_{\rm SFR}$(Z,z) shows a more pronounced low-metallicity tail at all redshifts
in the environment-dependent IMF scenario; the SFRD is less
concentrated in the most massive, metal rich galaxies than in the universal IMF case.
    \item The total SFRD is decreased with respect to the universal IMF case at all but
very low redshifts; the difference increases towards high redshifts and remains
within a factor of $\approx$2 for the considered IMF variations model.
    \item The assumption of the universal IMF may lead to an underestimate
of the fraction of NS and WD progenitors forming at low metallicity,
especially at low redshifts. 
It may also lead to an overestimate in the number of NS and WD progenitors forming
at $z\gtrsim 1$ and underestimate of the number of those stars forming at lower $z$.
The strength of this effect depends on the $f_{\rm SFR}$(Z,z).
The considered model for the IMF variations
has the strongest effect on the number and metallicity of 
formation of the WD progenitors with the masses $\gtrsim 1 M_{\odot}$,
relevant for the type Ia SN progenitors.
    \item The transition from the universal to environment dependent IMF has little
effect on the formation of the BH progenitors; the formation of the most massive
BH progenitors at low metallicity in the local Universe is slightly disfavored in the 
non-universal IMF scenario.
\end{enumerate}

\begin{acknowledgements}
We thank S{\o}ren Larsen and Jorryt Matthee for helpful discussions.
We thank the organisers and the participants of the workshop "Metals in Galaxies, Near and Far: Looking Ahead" and the Lorentz Center for an inspiring workshop that encouraged this study.
We thank the referee for their comments and a quick report.
MC and GN acknowledge support from the Netherlands Organisation for Scientific Research (NWO).
ZY acknowledges financial support from the China Scholarship Council (CSC, file number 201708080069). TJ acknowledges support by the Erasmus+ programme of the European Union under grant number 2017-1-CZ01- KA203-035562.
\end{acknowledgements}

\begin{appendix}

\section{Inferring iron abundances from the oxygen abundances}\label{app: FOHFeH}

The simplest assumption to compare the $Z_{O/H}$ and [Fe/H] metallicity measures would be that the [O/H] translates directly to [Fe/H] (maintaining the solar ratios; i.e. [O/H]=-1 translates to [Fe/H]=-1 etc.).
However, it may not be a very realistic one, as oxygen and iron are 
released to the interstellar medium on different timescales \citep[e.g.][]{WheelerSnedenTruran89}.
While the former is abundantly produced by massive stars on Myr timescales,
the latter is predominantly released in type Ia SN that show a much longer ($\approx$ 0.1 - 1 Gyr) delay with respect to the star formation episode.
Therefore, young, low--metallicity star--forming systems are generally expected to be
$\alpha$-enhanced \citep[i.e. show an overabundance of oxygen relative to iron with respect to solar, e.g.][]{ZhangZhao05,Izotov06}.
The difference in the timescales at which different elements are released to the interstellar medium is also reflected in the [O/Fe] vs [Fe/H] (or [O/H]) relation of a given stellar system:
the oldest and the most metal--poor stellar populations (with [Fe/H] $\lesssim$-1)
are generally $\alpha$-enhanced, with [O/Fe] approaching the ratio resulting from the core-collapse SN yields, but as the type Ia SNe start to release iron the relation bends, [O/Fe] decreases and gradually reaches the solar-scaled values.
However, the exact form of this relation (e.g. the location of the 'knee' and the extreme [O/Fe] values) depends on the SFH and the IMF of a particular stellar system and therefore is not universal \citep[e.g.][]{WheelerSnedenTruran89,Tolstoy09}.
For the purpose of this study we assume an example relation between [O/Fe] and [Fe/H]
that takes into account the overabundance of oxygen at low [Fe/H]:
\begin{equation}\label{eq: OFe-FeH}
    \mbox{[O/Fe]} = \begin{cases} 0.5 & \mbox{if [Fe/H]}<-1 \\ -0.5 \times \mbox{[Fe/H]} & \mbox{if [Fe/H]}\geqslant -1 \end{cases}
\end{equation}
which is loosely guided by the Milky Way relation \citep[e.g.][]{Bensby04,Reddy06,Bensby14,Steidel16}.
This relation is unconstrained at [Fe/H]>0.2 but up to this point shows 
no evidence for flattening  \citep[as is seen for other $\alpha$-elements, e.g.][]{Bensby04}. We note that whether we assume that there is a 
flattening above [Fe/H]>0.2 or extrapolate the relation from lower metallicities 
has no noticeable impact on our results.
The conversion between $Z_{\rm O/H}$ and [Fe/H] is then given by the relation:
\begin{equation}\label{eq: FOH-FeH}
 \mbox{[Fe/H]} = Z_{\rm O/H} - Z_{\rm O/H \odot} - \mbox{[O/Fe]}
\end{equation}
where $Z_{\rm O/H \odot}$=12 + log$_{10}$($\frac{\rm O_{\odot}}{\rm H_{\odot}}$) is the solar oxygen abundance. 
The resulting $Z_{\rm O/H}$--[Fe/H] relation is shown in Fig. \ref{fig: FOH-FeH_conversion}.

\section{The impact of the the IGIMF model and  metallicity conversion}\label{app: IGIMF}

\begin{figure}
    \centering
    \includegraphics[width=\hsize]{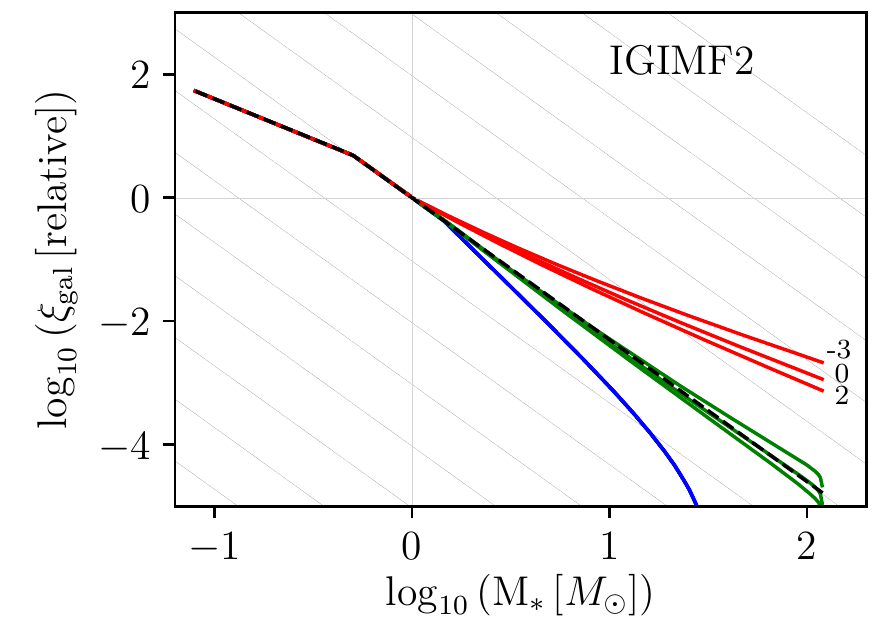}
    \caption{Galaxy-wide IMFs computed using IGIMF2 as a function of stellar mass plotted for several values of SFRs ($10^{-3}$, $1.0$, $10^{3}$ $M_{\odot}$/yr) and metallicities (-3, 0 2). The galaxy-wide IMFs are normalised by their values for 1 $M_{\odot}$ to show the slope changes. Note the main difference between the IGIMF2 and IGIMF3 (Fig. \ref{fig:IGIMF_3}) models are the variation for the low-mass part of the galaxy-wide IMF.}
    \label{fig:IGIMF_2}
\end{figure}

\begin{figure}
    \centering
    \includegraphics[width=\hsize]{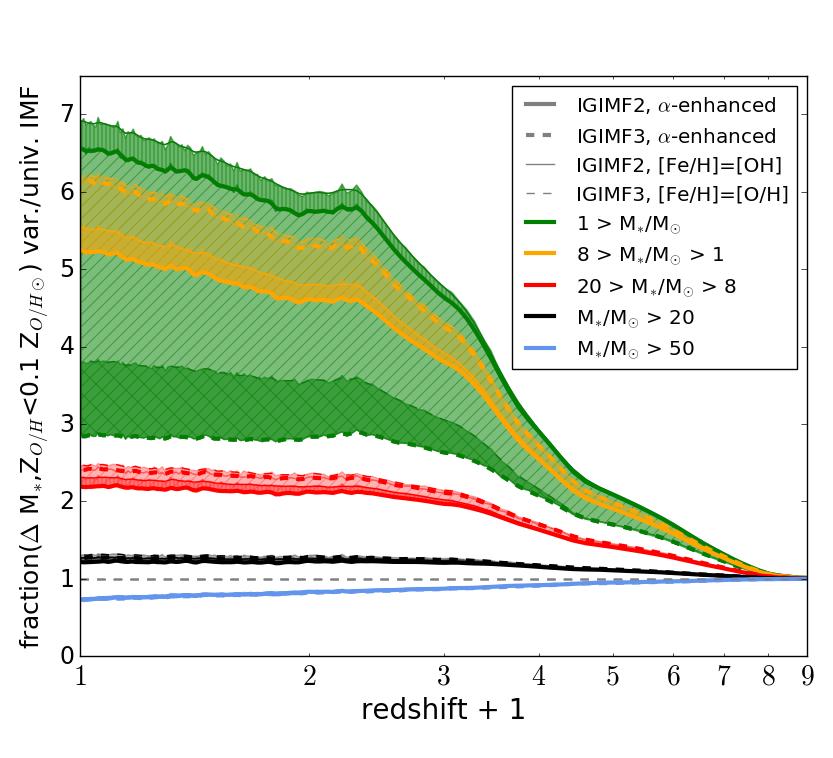}
    \caption{Ratio of the fraction of stars forming at low metallicity ($Z_{\rm O/H}<0.1 Z_{\rm O/H\odot}$) in different mass ranges in the case with the environment dependent IMF to the universal IMF as a function of redshift. The ratios are shown for the high metallicity extreme $f_{\rm SFR}$(Z,z) from \citet{ChruslinskaNelemans19} and for two non-universal IMF models from \citet{Jerabkova18}: IGIMF2 model - solid line; IGIMF3 model - dashed line. The large shaded area between the solid and dashed lines spans between the ratios obtained for these two IMF models. The ratios are also mildly affected by the conversion of 12 + log$_{10}$(O/H) to [Fe/H] metallicity measures. The thick solid and dashed curves were obtained for the default conversion described in Sec. \ref{sec: method}. Using conversion maintaining solar abundance ratios leads to slight increase in the estimated ratios (thin lines, additional shadings).}
    \label{fig: NlowZ_ratios_comparison}
\end{figure}

\begin{figure}
    \centering
    \includegraphics[width=\hsize]{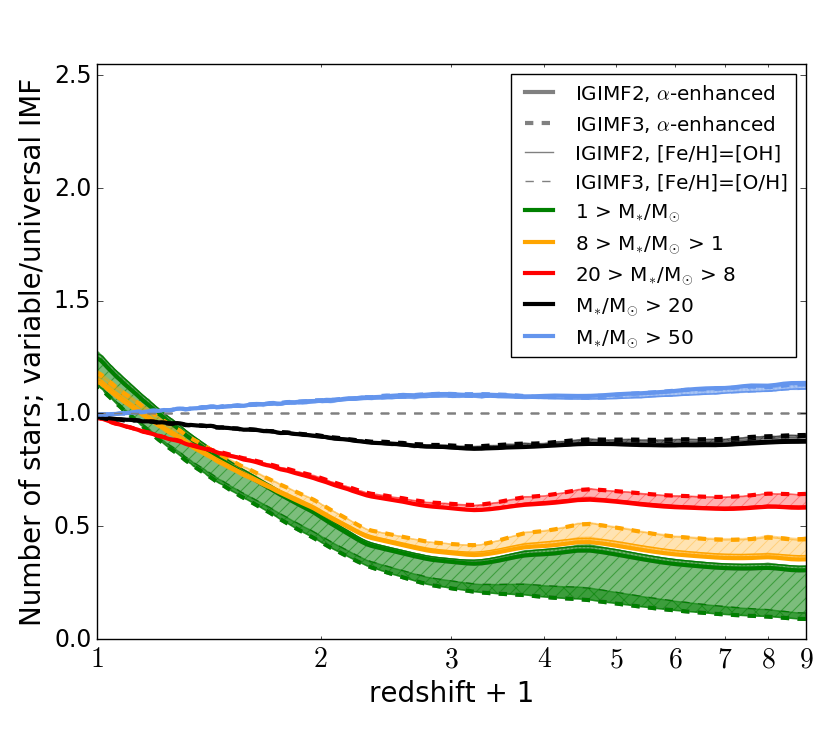}
    \caption{Ratio of the number of stars forming in different mass ranges in the case with the environment dependent IMF to the universal IMF as a function of redshift. The ratios are shown for the high metallicity extreme $f_{\rm SFR}$(Z,z) from \citet{ChruslinskaNelemans19} and for two non-universal IMF models from \citet{Jerabkova18}: IGIMF2 model - solid line; IGIMF3 model - dashed line. The large shaded area between the solid and dashed lines spans between the ratios obtained for these two IMF models. The ratios are also mildly affected by the conversion of the 12 + log$_{10}$(O/H) to [Fe/H] metallicity measures. The thick solid and dashed curves were obtained for the default conversion described in Sec. \ref{sec: method}. Using conversion maintaining solar abundance ratios leads to slight increase in the estimated ratios (thin lines, additional shadings).}
    \label{fig: N_ratios_comparison}
\end{figure}

The difference between the IGIMF2 and IGIMF3 models from \citet{Jerabkova18}
lies mainly in the description of the low mass part of the IMF ($M_{*}<1 M_{\odot}$), see Fig.~\ref{fig:IGIMF_2} and \ref{fig:IGIMF_3}.
In the IGIMF3 model the low mass end IMF slope is assumed to follow the tentative empirical metallicity
dependence described in \citet{Kroupa02} and \citet{Marks12} 
\footnote{ i.e.
$\alpha_{1,2} = \alpha_{\rm K01;1,2}$ + 0.5 [Fe/H], where $\alpha_{1}$ is the IMF slope between
0.08 - 0.5 $M_{\odot}$ with $\alpha_{\rm K01;1}$=1.3 and $\alpha_{2}$ between 0.5 - 1 $M_{\odot}$ with $\alpha_{\rm K01;2}$=2.3.
The relation was derived between [Fe/H] of -2 dex to 0.2 dex
and extrapolated outside this range}.
In the IGIMF2 prescription the low mass part of the IMF is fixed to the MW-like shape.
\\
In the IGIMF3 prescription, the IMF is increasingly bottom light at low metallicities,
so the formation efficiency of the low-mass stars ($<1 M_{\odot}$) is decreased.
The difference in the number of stars forming in this mass range in the IGIMF2 (IGIMF3)
scenario with respect to the universal IMF reaches to a factor of $\approx$3 ($\gtrsim$10)
at high redshifts (see Fig. \ref{fig: N_ratios_comparison}),
where the increasing fraction of the star formation happens at low metallicity.
When the IGIMF2 instead of the IGIMF3 description of the IMF variations is used,
the increase in the fraction of low-mass stars forming at low metallicity
with respect to the universal IMF case in the local Universe is enhanced by a factor of $\approx$1.5 - 2.5 (depending on the $f_{\rm SFR}$(Z,z) variation). 
This effect is the strongest for the high metallicity extreme and this case is shown in Figure \ref{fig: NlowZ_ratios_comparison}.
The shaded areas spanning between the solid and dashed lines in Fig. 
\ref{fig: NlowZ_ratios_comparison} correspond to the fractions obtained using the IGIMF2 and IGIMF3 models respectively.
It can be seen that the effect of the change between the two models on the results concerning more massive stars ($>1 M_{\odot}$) is much smaller.
\\
\newline
The effect of the choice of the IGIMF2/IGIMF3 model on the
obtained cosmic SFH is negligible (see Fig. \ref{fig: SFRD_comparison_appendix}).
This choice also affects the global $f_{\rm SFR}$(Z,z)
distribution (the fraction of stellar mass forming at low metallicity is increased in the IGIMF2 case with respect to the estimates based on the IGIMF3 model), but the effect is relatively small.
To quantify the differences between the different $f_{\rm SFR}$(Z,z) versions, \citetalias{ChruslinskaNelemans19} compare the fraction
of stellar mass that forms since $z=3$ at low ($<0.1 Z_{\rm O/H \odot}$) and high ($>Z_{\rm O/H \odot}$) metallicity.
We note that those quantities change by less than 2\%
between the $f_{\rm SFR}$(Z,z) obtained using IGIMF2 and IGIMF3 models
(for a certain input universal IMF-based $f_{\rm SFR}$(Z,z)).
For a fixed IMF model, that difference is at the level of 1\% depending on the adopted metallicity conversion.
For comparison, the low and high metallicity extremes
are found to differ by 18\% (22\%) in terms of the low metallicity
mass fraction and by 26\% (20\%) in terms of the high metallicity
mass fraction for the universal (non-universal) IMF scenario.
\newline
\\
The additional shaded areas above the thick solid
and dashed lines in Fig. \ref{fig: NlowZ_ratios_comparison}
and Fig. \ref{fig: N_ratios_comparison}
indicate the fractions that would be obtained
if $Z_{\rm O/H}$ was assumed to translate to [Fe/H]
maintaining the solar abundance ratios, instead
of following the conversion described in Sec. \ref{sec: method}. 
See Sec. \ref{sec: discussion: metallicity} for discussion.

\section{The impact of the $f_{\rm SFR}$(Z,z)}\label{app: SFRD_Z_z}

The key difference between low and high-Z extreme $f_{\rm SFR}$(Z,z)
that is responsible for the differences in the ratios shown in Fig.
\ref{fig: N_ratios_IGIMF3} lies in the adopted SFR -- (galaxy) mass relation.
In the low-Z extreme the relation is assumed to flatten at high masses,
so the most massive galaxies have much lower SFR than in the high-Z extreme
(up to a factor of $\approx$100 for the highest masses, see Fig. 5 in \citetalias{ChruslinskaNelemans19}).
In the low redshift range this means that all galaxies in the low-Z extreme
form stars at higher rates than expected in the universal IMF case
(i.e. all receive positive corrections, see Fig. \ref{fig: SFR-corr factor}),
and hence the number of stars forming in different mass ranges
is generally higher in the non-universal than in the universal IMF case.
\\
The SFR of the highly star forming galaxies in the high-Z extreme is decreased with respect to the universal IMF case even at low redshifts,
which results in smaller (or only slightly increased at low redshifts) number of stars forming in the non-universal IMF case with respect to the universal IMF scenario.
\\
The difference in the metallicity description in the low and high-Z extreme
becomes important for the results shown in Fig.
\ref{fig: N_ratios_IGIMF3} only for the lowest considered mass range,
due to strong dependence of the low mass part of the IMF on metallicity.
In the adopted IMF variations model (IGIMF3 from \citet{Jerabkova18} -- this difference does not have a significant effect if the IGIMF2 model is used instead - see App. \ref{app: IGIMF}),
the formation of low-mass stars is favoured at high metallicities
and this counteracts the effect of the SFR corrections in the metal rich, 
highly star forming galaxies on the formation of stars $<1 M_{\odot}$, leading to relatively small difference in the ratios obtained for the low and high-Z extreme in this mass range.

\subsection{The discrepancy at high z reported in ChN19 and the non-universal IMF}\label{app: discrepancy}

\begin{figure}
    \centering
    \includegraphics[width=\hsize]{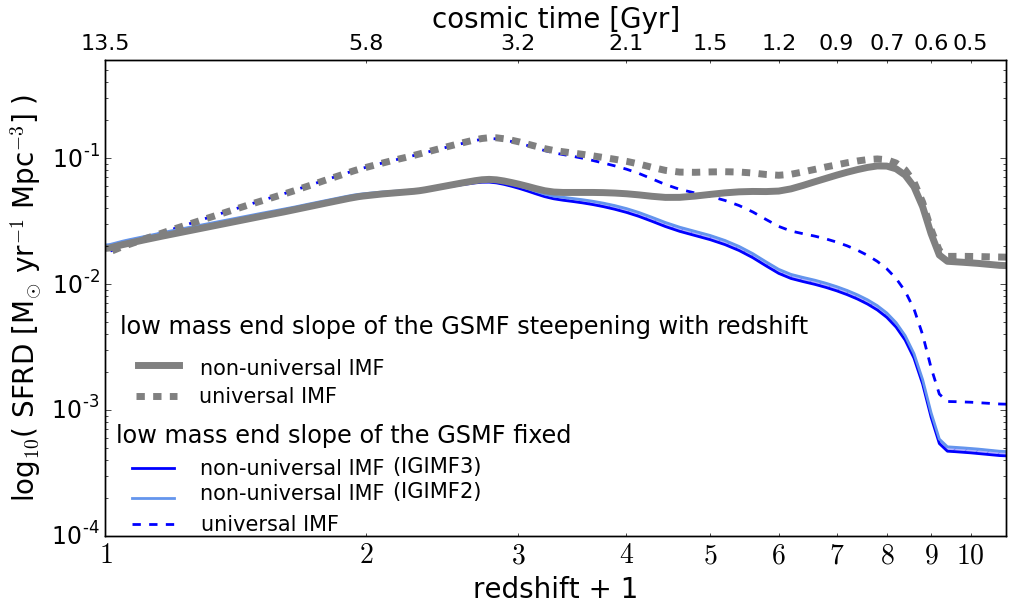}
    \caption{Cosmic SFH plotted for the universal (dashed lines) and non-universal (solid lines) IMF case using the high-Z extreme $f_{\rm SFR}$(Z,z) (thin lines) and the $f_{\rm SFR}$(Z,z) with the same MZR and SFMR, but with the low mass end slope of the GSMF allowed to steepen with redshift as discussed in \citetalias{ChruslinskaNelemans19} (thick lines). For the high-Z extreme example we also show the the cosmic SFH obtained using the IGIMF2 (light blue) or IGIMF3 (dark blue) non-universal IMF model. The former is slightly higher, but the difference is negligible.}
    \label{fig: SFRD_comparison_appendix}
\end{figure}

\begin{figure}
    \centering
    \includegraphics[width=\hsize]{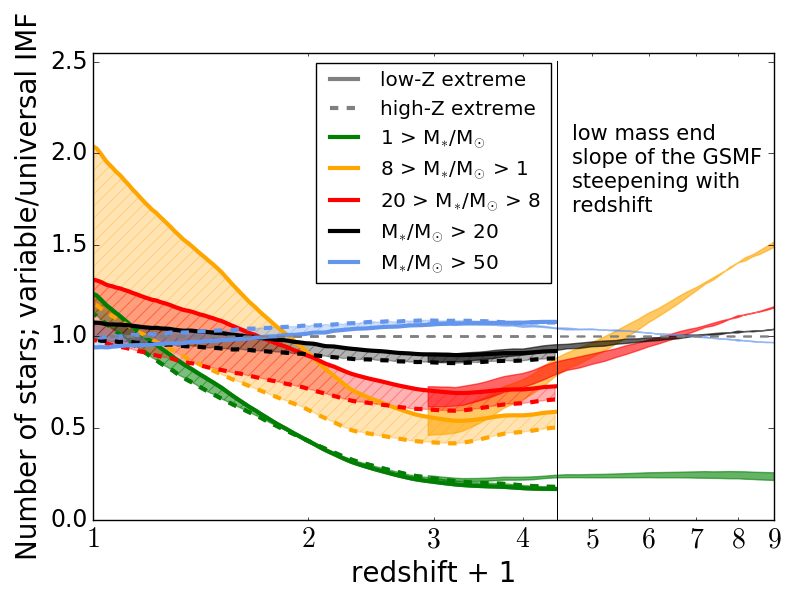}
    \caption{Same as Fig. \ref{fig: N_ratios_IGIMF3}, but with additional lines included to demonstrate the impact of the assumption about the low mass end slope of the GSMF. At $z\lesssim 2$ the ratio is not affected by this assumption. At $z\gtrsim 3$, the difference becomes apparent -- if the low mass end slope of the GSMF is fixed (as in the cases shown in Fig. \ref{fig: N_ratios_IGIMF3}), the trend flattens and the ratio stays at the level reached at $z\sim 3$. If the slope steepens with redshift, the ratio for the stars in the intermediate-mass WD and NS progenitors mass range starts increasing with $z$.}
    \label{fig: Nratio_GSMF_appendix}
\end{figure}

The low and high metallicity extremes as discussed above 
and referred to throughout this paper were obtained in \citetalias{ChruslinskaNelemans19} under the
assumption of a non-evolving low mass end slope of the galaxy
stellar mass function. As discussed in \citetalias{ChruslinskaNelemans19},
there are some observational indications of the steepening trend
of this slope with increasing redshift (see Sec. 3.1 therein).
This leads to an increasing number density of the low-mass galaxies, and 
hence the increasing contribution of the star formation at low metallicity
to the total SFRD budget at high $z$.
\\
This assumption has little effect on the results at $z\lesssim$3.
However, it becomes important at higher $z$:  
if the low mass end slope of the GSMF is allowed to steepen
with redshift (and all the relations are simply extrapolated down
to $M_{\rm gal,*}=10^{6} M_{\odot}$), the $z\gtrsim5$ $f_{\rm SFR}$(Z,z) 
is completely dominated
by the low-mass galaxies and low-metallicity star formation.
\citetalias{ChruslinskaNelemans19} consider such variations of the 
$f_{\rm SFR}$(Z,z) at high $z$ unrealistic, as they lead to 
the flattening, or even a secondary peak at $z\sim7$
in the cosmic SFRD,
instead of the observed decrease at $z\gtrsim 2$.
This leads to a significant
overestimate in the total SFRD at these high redshifts with respect
to other observational estimates of the cosmic SFRD
(see Sec. 4.1.1 and 6.1 therein).
\\
As shown in Sec. \ref{sec: results}, the SFRD at high $z$ 
in the low and high-Z extremes $f_{\rm SFR}$(Z,z) is lower in the
non-universal IMF than in the universal IMF scenario.
However, the discrepancy reported in \citetalias{ChruslinskaNelemans19}
in the $f_{\rm SFR}$(Z,z) cases with steepening low mass end of the GSMF
is due to increasing contribution of the galaxies of the lowest masses
and SFRs. The SFR of those galaxies in the considered non-universal
IMF scenario would be increased due to positive corrections even at high $z$
(see Fig. \ref{fig: IGIMF3}).
The lowered SFR of the more massive galaxies in the non-universal IMF scenario
is not enough to compensate for that effect and
the high-$z$ SFRD still shows a peak at high $z$,
which in the non-universal IMF scenario is higher than the peak at $z\sim 2$.
This is shown in Fig. \ref{fig: SFRD_comparison_appendix} 
for the high-Z extreme $f_{\rm SFR}$(Z,z) example.
\\
We note that the direct comparison of the results discussed in this study
with the observational cosmic SFRD estimates \citep[e.g.][]{MadauDickinson14,Bouwens15,MadauFragos17, Fermi18}
as done in \citetalias{ChruslinskaNelemans19}
is no longer possible, as those estimates assume the universal IMF.
\\
\newline
The assumption about the low mass end slope of the GSMF
would also affect the results shown in Fig. \ref{fig: N_ratios_IGIMF3} and Fig. \ref{fig: NlowZ_ratios_IGIMF3} at high redshifts. 
The fraction of the number of stars forming
in the non-universal versus universal IMF scenario for different
assumptions about the low mass end of the GSMF
is demonstrated in Fig. \ref{fig: Nratio_GSMF_appendix}.
If the low mass end of the GSMF steepens with redshift,
the number of stars forming at $z\gtrsim 3$
in the massive WD progenitor and NS and lower mass BH progenitor
mass range is increased in the non-universal with respect
to universal IMF scenario.
In that case the star formation at high $z$ happens
predominantly in galaxies 
with low metallicity and SFR $\approx 0.1 M_{\odot}/yr$
that in the adopted non-universal IMF scenario 
have simultaneously bottom and top light IMFs,
which favours the formation of stars in this mass range
with respect to universal IMF.
The main difference in the ratio of the 
fraction of stars forming at low metallicity in the
non-universal versus universal IMF case 
due to the GSMF with the steepening slope
would be an earlier decrease at $z>2$
(as the fraction of stars forming at low Z approaches
1 at lower redshift in this GSMF variation).
  
 \section{Zoom into the type Ia SN progenitor mass range}\label{app: SNIa}
 
\begin{figure}
    \centering
    \includegraphics[width=\hsize]{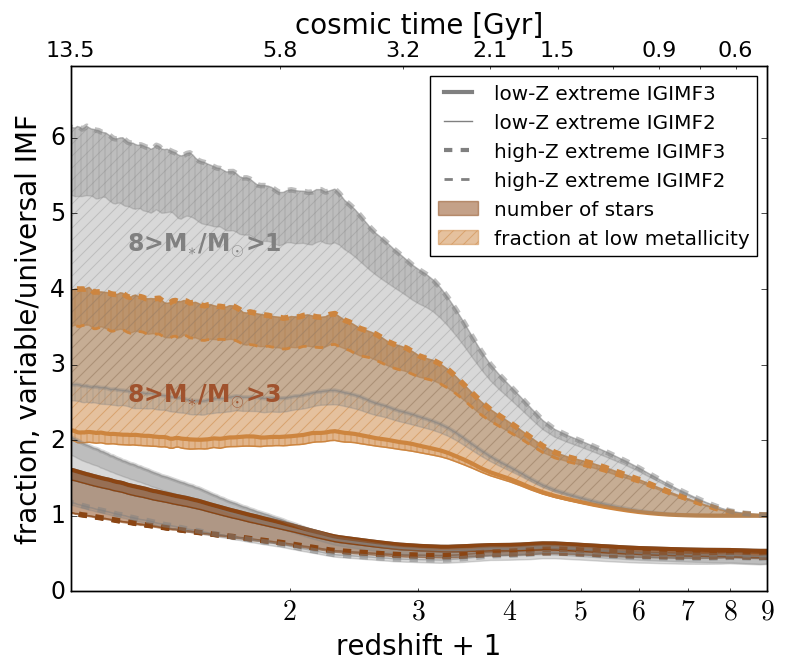}
    \caption{Ratio of the fraction of stars forming at low metallicity ($Z_{\rm O/H}<Z_{\rm O/H\odot}$; hatched filled area) and the ratio of the number of stars (filled area) in the case with the environment dependent IMF to the universal IMF as a function of redshift. The solid lines correspond to low-Z extreme and the dashed lines to high-Z extreme $f_{\rm SFR}$(Z,z) distributions. The thick lines correspond to IGIMF3 model, while the thin lines to the IGIMF2 (lower than the corresponding IGIMF3 lines). The coloured areas correspond to stars with the initial masses between 3 - 8 $M_{\odot}$ - i.e. the mass range relevant for the formation of the progenitors of the type Ia supernovae. The grey areas in the background correspond to the WD progenitors with initial masses in the range 1 - 8 $M_{\odot}$ (same as in Fig. \ref{fig: N_ratios_IGIMF3}-\ref{fig: NlowZ_ratios_IGIMF3}).}
    \label{fig: ratio_type_Ia}
\end{figure}

 Figure \ref{fig: ratio_type_Ia} shows the fraction of 
 low-metallicity stars and the
 number of stars forming between 3 - 8 $\Msun$ mass range
 in the non-universal and universal IMF scenario.
 The additional shadings demonstrate the difference between the
 IGIMF2 and IGIMF3 (Fig.~\ref{fig: IGIMF3}) non-universal IMF models.
 
\end{appendix}
%
%

\bibliographystyle{aa}
\bibliography{bib_SNnotes.bib}

\begin{thebibliography}{102}
\expandafter\ifx\csname natexlab\endcsname\relax\def\natexlab#1{#1}\fi

\bibitem[{{Adams} \& {Fatuzzo}(1996)}]{Adams_Fatuzzo1996}
{Adams}, F.~C. \& {Fatuzzo}, M. 1996, \apj, 464, 256

\bibitem[{{Adams} \& {Laughlin}(1996)}]{Adams_Laughlin1996}
{Adams}, F.~C. \& {Laughlin}, G. 1996, \apj, 468, 586

\bibitem[{{Anders} \& {Grevesse}(1989)}]{AndersGrevesse89}
{Anders}, E. \& {Grevesse}, N. 1989, \gca, 53, 197

\bibitem[{{Asplund} {et~al.}(2009){Asplund}, {Grevesse}, {Sauval}, \&
  {Scott}}]{Asplund09}
{Asplund}, M., {Grevesse}, N., {Sauval}, A.~J., \& {Scott}, P. 2009, \araa, 47,
  481

\bibitem[{{Belczynski} {et~al.}(2010){Belczynski}, {Bulik}, {Fryer}, {Ruiter},
  {Valsecchi}, {Vink}, \& {Hurley}}]{Belczynski10Mmax}
{Belczynski}, K., {Bulik}, T., {Fryer}, C.~L., {et~al.} 2010, \apj, 714, 1217

\bibitem[{{Bensby} {et~al.}(2004){Bensby}, {Feltzing}, \&
  {Lundstr{\"o}m}}]{Bensby04}
{Bensby}, T., {Feltzing}, S., \& {Lundstr{\"o}m}, I. 2004, \aap, 415, 155

\bibitem[{{Bensby} {et~al.}(2014){Bensby}, {Feltzing}, \& {Oey}}]{Bensby14}
{Bensby}, T., {Feltzing}, S., \& {Oey}, M.~S. 2014, \aap, 562, A71

\bibitem[{{Boogaard} {et~al.}(2018){Boogaard}, {Brinchmann}, {Bouch{\'e}},
  {Paalvast}, {Bacon}, {Bouwens}, {Contini}, {Gunawardhana}, {Inami}, {Marino},
  {Maseda}, {Mitchell}, {Nanayakkara}, {Richard}, {Schaye}, {Schreiber},
  {Tacchella}, {Wisotzki}, \& {Zabl}}]{Boogaard18}
{Boogaard}, L.~A., {Brinchmann}, J., {Bouch{\'e}}, N., {et~al.} 2018, \aap,
  619, A27

\bibitem[{{Bouwens} {et~al.}(2015){Bouwens}, {Illingworth}, {Oesch}, {Trenti},
  {Labb{\'e}}, {Bradley}, {Carollo}, {van Dokkum}, {Gonzalez}, {Holwerda},
  {Franx}, {Spitler}, {Smit}, \& {Magee}}]{Bouwens15}
{Bouwens}, R.~J., {Illingworth}, G.~D., {Oesch}, P.~A., {et~al.} 2015, \apj,
  803, 34

\bibitem[{{Brown} \& {Wilson}(2019)}]{Brown2019}
{Brown}, T. \& {Wilson}, C.~D. 2019, \apj, 879, 17

\bibitem[{{Bruzual} \& {Charlot}(2003)}]{BruzualChalot03}
{Bruzual}, G. \& {Charlot}, S. 2003, \mnras, 344, 1000

\bibitem[{{Chruslinska} \& {Nelemans}(2019)}]{ChruslinskaNelemans19}
{Chruslinska}, M. \& {Nelemans}, G. 2019, \mnras, 488, 5300

\bibitem[{{Chruslinska} {et~al.}(2019){Chruslinska}, {Nelemans}, \&
  {Belczynski}}]{Chruslinska19}
{Chruslinska}, M., {Nelemans}, G., \& {Belczynski}, K. 2019, \mnras, 482, 5012

\bibitem[{{Claeys} {et~al.}(2014){Claeys}, {Pols}, {Izzard}, {Vink}, \&
  {Verbunt}}]{Clays14}
{Claeys}, J.~S.~W., {Pols}, O.~R., {Izzard}, R.~G., {Vink}, J., \& {Verbunt},
  F.~W.~M. 2014, \aap, 563, A83

\bibitem[{{Conroy} {et~al.}(2017){Conroy}, {van Dokkum}, \&
  {Villaume}}]{Conroy2017}
{Conroy}, C., {van Dokkum}, P.~G., \& {Villaume}, A. 2017, \apj, 837, 166

\bibitem[{{Conselice} {et~al.}(2016){Conselice}, {Wilkinson}, {Duncan}, \&
  {Mortlock}}]{Conselice16}
{Conselice}, C.~J., {Wilkinson}, A., {Duncan}, K., \& {Mortlock}, A. 2016,
  \apj, 830, 83

\bibitem[{{Cooper} {et~al.}(2009){Cooper}, {Newman}, \& {Yan}}]{Cooper09}
{Cooper}, M.~C., {Newman}, J.~A., \& {Yan}, R. 2009, \apj, 704, 687

\bibitem[{Dabringhausen {et~al.}(2009)Dabringhausen, Kroupa, \&
  Baumgardt}]{Dabringhausen2009}
Dabringhausen, J., Kroupa, P., \& Baumgardt, H. 2009, \mnras, 394, 1529

\bibitem[{{Dabringhausen} {et~al.}(2012){Dabringhausen}, {Kroupa}, \&
  {Pflamm-Altenburg}}]{Dabringhausen2012}
{Dabringhausen}, J., {Kroupa}, P., \& {Pflamm-Altenburg}, J. and.~{Mieske}, S.
  2012, \apj, 747, 72

\bibitem[{{de Mink} {et~al.}(2013){de Mink}, {Langer}, {Izzard}, {Sana}, \& {de
  Koter}}]{deMink13}
{de Mink}, S.~E., {Langer}, N., {Izzard}, R.~G., {Sana}, H., \& {de Koter}, A.
  2013, \apj, 764, 166

\bibitem[{{Delahaye} \& {Pinsonneault}(2006)}]{DelahayePinsonneault06}
{Delahaye}, F. \& {Pinsonneault}, M.~H. 2006, \apj, 649, 529

\bibitem[{{Dib} {et~al.}(2007){Dib}, {Kim}, \& {Shadmehri}}]{Dib2007}
{Dib}, S., {Kim}, J., \& {Shadmehri}, M. 2007, \mnras, 381, L40

\bibitem[{{Doherty} {et~al.}(2015){Doherty}, {Gil-Pons}, {Siess}, {Lattanzio},
  \& {Lau}}]{Doherty15}
{Doherty}, C.~L., {Gil-Pons}, P., {Siess}, L., {Lattanzio}, J.~C., \& {Lau}, H.
  H.~B. 2015, \mnras, 446, 2599

\bibitem[{{Ekstr{\"o}m} {et~al.}(2012){Ekstr{\"o}m}, {Georgy}, {Eggenberger},
  {Meynet}, {Mowlavi}, {Wyttenbach}, {Granada}, {Decressin}, {Hirschi},
  {Frischknecht}, {Charbonnel}, \& {Maeder}}]{Ekstrom12}
{Ekstr{\"o}m}, S., {Georgy}, C., {Eggenberger}, P., {et~al.} 2012, \aap, 537,
  A146

\bibitem[{{Eldridge} \& {Stanway}(2009)}]{Eldridge2009}
{Eldridge}, J.~J. \& {Stanway}, E.~R. 2009, \mnras, 400, 1019

\bibitem[{{Eldridge} {et~al.}(2017){Eldridge}, {Stanway}, {Xiao}, {McClelland
  }, {Taylor}, {Ng}, {Greis}, \& {Bray}}]{Eldridge2017}
{Eldridge}, J.~J., {Stanway}, E.~R., {Xiao}, L., {et~al.} 2017, \pasa, 34, e058

\bibitem[{{Fermi-LAT Collaboration} {et~al.}(2018){Fermi-LAT Collaboration},
  {Abdollahi}, {Ackermann}, {Ajello}, {Atwood}, {Baldini}, {Ballet},
  {Barbiellini}, {Bastieri}, {Becerra Gonzalez}, {Bellazzini}, {Bissaldi},
  {Blandford}, {Bloom}, {Bonino}, {Bottacini}, {Buson}, {Bregeon}, {Bruel},
  {Buehler}, {Cameron}, {Caputo}, {Caraveo}, {Cavazzuti}, {Charles}, {Chen},
  {Cheung}, {Chiaro}, {Ciprini}, {Cohen-Tanugi}, {Cominsky}, {Conrad},
  {Costantin}, {Cutini}, {D'Ammando}, {de Palma}, {Desai}, {Digel}, {Di Lalla},
  {Di Mauro}, {Di Venere}, {Dom{\'\i}nguez}, {Favuzzi}, {Fegan}, {Finke},
  {Franckowiak}, {Fukazawa}, {Funk}, {Fusco}, {Gallardo Romero}, {Gargano},
  {Gasparrini}, {Giglietto}, {Giordano}, {Giroletti}, {Green}, {Grenier},
  {Guillemot}, {Guiriec}, {Hartmann}, {Hays}, {Helgason}, {Horan},
  {J{\'o}hannesson}, {Kocevski}, {Kuss}, {Larsson}, {Latronico}, {Li}, {Longo},
  {Loparco}, {Lott}, {Lovellette}, {Lubrano}, {Madejski}, {Magill}, {Maldera},
  {Manfreda}, {Marcotulli}, {Mazziotta}, {McEnery}, {Meyer}, {Michelson},
  {Mizuno}, {Monzani}, {Morselli}, {Moskalenko}, {Negro}, {Nuss}, {Ojha},
  {Omodei}, {Orienti}, {Orlando}, {Ormes}, {Palatiello}, {Paliya}, {Paneque},
  {Perkins}, {Persic}, {Pesce-Rollins}, {Petrosian}, {Piron}, {Porter},
  {Primack}, {Principe}, {Rain{\`o}}, {Rando}, {Razzano}, {Razzaque}, {Reimer},
  {Reimer}, {Saz Parkinson}, {Sgr{\`o}}, {Siskind}, {Spandre}, {Spinelli},
  {Suson}, {Tajima}, {Takahashi}, {Thayer}, {Tibaldo}, {Torres}, {Torresi},
  {Tosti}, {Tramacere}, {Troja}, {Valverde}, {Vianello}, {Vogel}, {Wood}, \&
  {Zaharijas}}]{Fermi18}
{Fermi-LAT Collaboration}, {Abdollahi}, S., {Ackermann}, M., {et~al.} 2018,
  Science, 362, 1031

\bibitem[{{Ferreras} {et~al.}(2015){Ferreras}, {Weidner}, {Vazdekis}, \& {La
  Barbera}}]{Ferraras2015}
{Ferreras}, I., {Weidner}, C., {Vazdekis}, A., \& {La Barbera}, F. 2015,
  \mnras, 448, L82

\bibitem[{{Fioc} {et~al.}(2011){Fioc}, {Le Borgne}, \&
  {Rocca-Volmerange}}]{Fioc2011}
{Fioc}, M., {Le Borgne}, D., \& {Rocca-Volmerange}, B. 2011, {P{\'E}GASE:
  Metallicity-consistent Spectral Evolution Model of Galaxies}, Astrophysics
  Source Code Library

\bibitem[{{Fioc} \& {Rocca-Volmerange}(1999)}]{Fioc1999}
{Fioc}, M. \& {Rocca-Volmerange}, B. 1999, ArXiv Astrophysics e-prints
  [\eprint{astro-ph/9912179}], code description, only astro-ph version:
  astro-ph/9912179

\bibitem[{{Fioc} \& {Rocca-Volmerange}(2019)}]{Fioc19}
{Fioc}, M. \& {Rocca-Volmerange}, B. 2019, \aap, 623, A143

\bibitem[{{Fontanot} {et~al.}(2017){Fontanot}, {De Lucia}, {Hirschmann},
  {Bruzual}, \& {Zibetti}}]{Fontanot2017}
{Fontanot}, F., {De Lucia}, G., {Hirschmann}, M., {Bruzual}, G. and.~{Charlot},
  S., \& {Zibetti}, S. 2017, \mnras, 464, 3812

\bibitem[{{Gargiulo} {et~al.}(2015){Gargiulo}, {Cora}, {Padilla}, {Mu{\~n}oz
  Arancibia}, {Ruiz}, {Orsi}, {Tecce}, {Weidner}, \& {Bruzual}}]{Gargiulo2015}
{Gargiulo}, I.~D., {Cora}, S.~A., {Padilla}, N.~D., {et~al.} 2015, \mnras, 446,
  3820

\bibitem[{{G{\"o}tberg} {et~al.}(2019){G{\"o}tberg}, {de Mink}, {Groh},
  {Leitherer}, \& {Norman}}]{Gotberg19}
{G{\"o}tberg}, Y., {de Mink}, S.~E., {Groh}, J.~H., {Leitherer}, C., \&
  {Norman}, C. 2019, \aap, 629, A134

\bibitem[{{Heger} {et~al.}(2003{\natexlab{a}}){Heger}, {Fryer}, {Woosley},
  {Langer}, \& {Hartmann}}]{Heger03}
{Heger}, A., {Fryer}, C.~L., {Woosley}, S.~E., {Langer}, N., \& {Hartmann},
  D.~H. 2003{\natexlab{a}}, \apj, 591, 288

\bibitem[{{Heger} {et~al.}(2003{\natexlab{b}}){Heger}, {Fryer}, {Woosley},
  {Langer}, \& {Hartmann}}]{Heger2003}
{Heger}, A., {Fryer}, C.~L., {Woosley}, S.~E., {Langer}, N., \& {Hartmann},
  D.~H. 2003{\natexlab{b}}, \apj, 591, 288

\bibitem[{{Hopkins}(2018)}]{Hopkins18}
{Hopkins}, A.~M. 2018, \pasa, 35, 39

\bibitem[{{Ibeling} \& {Heger}(2013)}]{Ibeling13}
{Ibeling}, D. \& {Heger}, A. 2013, \apjl, 765, L43

\bibitem[{{Izotov} {et~al.}(2006){Izotov}, {Stasi{\'n}ska}, {Meynet}, {Guseva},
  \& {Thuan}}]{Izotov06}
{Izotov}, Y.~I., {Stasi{\'n}ska}, G., {Meynet}, G., {Guseva}, N.~G., \&
  {Thuan}, T.~X. 2006, \aap, 448, 955

\bibitem[{{Je{\v{r}}{\'a}bkov{\'a}} {et~al.}(2018){Je{\v{r}}{\'a}bkov{\'a}},
  {Hasani Zonoozi}, {Kroupa}, {Beccari}, {Yan}, {Vazdekis}, \&
  {Zhang}}]{Jerabkova18}
{Je{\v{r}}{\'a}bkov{\'a}}, T., {Hasani Zonoozi}, A., {Kroupa}, P., {et~al.}
  2018, \aap, 620, A39

\bibitem[{{Je{\v{r}}{\'a}bkov{\'a}} {et~al.}(2017){Je{\v{r}}{\'a}bkov{\'a}},
  {Kroupa}, {Dabringhausen}, {Hilker}, \& {Bekki}}]{Jerabkova17}
{Je{\v{r}}{\'a}bkov{\'a}}, T., {Kroupa}, P., {Dabringhausen}, J., {Hilker}, M.,
  \& {Bekki}, K. 2017, \aap, 608, A53

\bibitem[{{Joncour} {et~al.}(2018){Joncour}, {Duch{\^e}ne}, {Moraux and}, \&
  {Motte}}]{Joncour2018}
{Joncour}, I., {Duch{\^e}ne}, G., {Moraux and}, E., \& {Motte}, F. 2018, ArXiv
  e-prints [\eprint[arXiv]{1809.02380}], accepted to A\&A

\bibitem[{{Kalari} {et~al.}(2018){Kalari}, {Carraro}, {Evans}, \&
  {Rubio}}]{Kalari2018}
{Kalari}, V.~M., {Carraro}, G., {Evans}, C.~J., \& {Rubio}, M. 2018, \apj, 857,
  132

\bibitem[{{Kennicutt}(1998)}]{Kennicutt98}
{Kennicutt}, Robert~C., J. 1998, \araa, 36, 189

\bibitem[{{Kennicutt} \& {Evans}(2012)}]{KennicuttEvans12}
{Kennicutt}, R.~C. \& {Evans}, N.~J. 2012, \araa, 50, 531

\bibitem[{{Kroupa}(2001)}]{Kroupa01}
{Kroupa}, P. 2001, \mnras, 322, 231

\bibitem[{{Kroupa}(2002{\natexlab{a}})}]{Kroupa2002}
{Kroupa}, P. 2002{\natexlab{a}}, Science, 295, 82

\bibitem[{{Kroupa}(2002{\natexlab{b}})}]{Kroupa02}
{Kroupa}, P. 2002{\natexlab{b}}, Science, 295, 82

\bibitem[{{Kroupa}(2005)}]{Kroupa2005}
{Kroupa}, P. 2005, in ESA Special Publication, Vol. 576, The Three-Dimensional
  Universe with Gaia, ed. C.~{Turon}, K.~S. {O'Flaherty}, \& M.~A.~C.
  {Perryman}, 629

\bibitem[{{Kroupa} \& {Weidner}(2003)}]{KroupaWeidner03}
{Kroupa}, P. \& {Weidner}, C. 2003, \apj, 598, 1076

\bibitem[{{Kroupa} {et~al.}(2013){Kroupa}, {Weidner}, {Pflamm-Altenburg},
  {Thies}, {Dabringhausen}, {Marks}, \& {Maschberger}}]{Kroupa13}
{Kroupa}, P., {Weidner}, C., {Pflamm-Altenburg}, J., {et~al.} 2013, {The
  Stellar and Sub-Stellar Initial Mass Function of Simple and Composite
  Populations}, ed. T.~D. {Oswalt} \& G.~{Gilmore}, Vol.~5, 115

\bibitem[{{Lada} \& {Lada}(2003)}]{lada2003}
{Lada}, C.~J. \& {Lada}, E.~A. 2003, \araa, 41, 57

\bibitem[{{Larson}(1998)}]{Larson1998}
{Larson}, R.~B. 1998, \mnras, 301, 569

\bibitem[{{Lee} {et~al.}(2009){Lee}, {Gil de Paz}, {Tremonti}, {Kennicutt},
  {Bothwell}, {Calzetti}, {Dalcanton}, {Engelbracht}, {Funes}, {Sakai},
  {Skillman}, {van Zee}, \& {Weisz}}]{Lee2009}
{Lee}, J.~C., {Gil de Paz}, A., {Tremonti}, C., {et~al.} 2009, \apj, 706, 599

\bibitem[{{Lee} {et~al.}(2015){Lee}, {Sanders}, {Casey}, {Toft}, {Scoville},
  {Hung}, {Le Floc'h}, {Ilbert}, {Zahid}, {Aussel}, {Capak}, {Kartaltepe},
  {Kewley}, {Li}, {Schawinski}, {Sheth}, \& {Xiao}}]{Lee15}
{Lee}, N., {Sanders}, D.~B., {Casey}, C.~M., {et~al.} 2015, \apj, 801, 80

\bibitem[{{Leitherer} {et~al.}(1999){Leitherer}, {Schaerer}, {Goldader},
  {Delgado}, {Robert}, {Kune}, {de Mello}, {Devost}, \&
  {Heckman}}]{Leitherer1999}
{Leitherer}, C., {Schaerer}, D., {Goldader}, J.~D., {et~al.} 1999, \apjs, 123,
  3

\bibitem[{{Livio} \& {Mazzali}(2018)}]{LivioMazzali18}
{Livio}, M. \& {Mazzali}, P. 2018, \physrep, 736, 1

\bibitem[{{Madau} \& {Dickinson}(2014)}]{MadauDickinson14}
{Madau}, P. \& {Dickinson}, M. 2014, \araa, 52, 415

\bibitem[{{Madau} \& {Fragos}(2017)}]{MadauFragos17}
{Madau}, P. \& {Fragos}, T. 2017, \apj, 840, 39

\bibitem[{{Maiolino} \& {Mannucci}(2019)}]{MaiolinoMannucci19}
{Maiolino}, R. \& {Mannucci}, F. 2019, \aapr, 27, 3

\bibitem[{{Mandel} \& {Farmer}(2018)}]{MandelFarmer18}
{Mandel}, I. \& {Farmer}, A. 2018, arXiv e-prints, arXiv:1806.05820

\bibitem[{{Mannucci} {et~al.}(2010){Mannucci}, {Cresci}, {Maiolino}, {Marconi},
  \& {Gnerucci}}]{Mannucci10}
{Mannucci}, F., {Cresci}, G., {Maiolino}, R., {Marconi}, A., \& {Gnerucci}, A.
  2010, \mnras, 408, 2115

\bibitem[{{Maoz} \& {Mannucci}(2012)}]{MaozMannucci12}
{Maoz}, D. \& {Mannucci}, F. 2012, \pasa, 29, 447

\bibitem[{Maoz {et~al.}(2014)Maoz, Mannucci, \& Nelemans}]{Maoz14}
Maoz, D., Mannucci, F., \& Nelemans, G. 2014, Annual Review of Astronomy and
  Astrophysics, 52, 107

\bibitem[{{Marks} {et~al.}(2012{\natexlab{a}}){Marks}, {Kroupa},
  {Dabringhausen}, \& {Pawlowski}}]{Marks2012TH}
{Marks}, M., {Kroupa}, P., {Dabringhausen}, J., \& {Pawlowski}, M.~S.
  2012{\natexlab{a}}, \mnras, 422, 2246

\bibitem[{{Marks} {et~al.}(2012{\natexlab{b}}){Marks}, {Kroupa},
  {Dabringhausen}, \& {Pawlowski}}]{Marks12}
{Marks}, M., {Kroupa}, P., {Dabringhausen}, J., \& {Pawlowski}, M.~S.
  2012{\natexlab{b}}, \mnras, 422, 2246

\bibitem[{{Mart{\'{\i}}n-Navarro} {et~al.}(2015){Mart{\'{\i}}n-Navarro}, {La
  Barbera}, {Vazdekis}, {Falc{\'o}n-Barroso}, \& {Ferreras}}]{Martin-Navarro15}
{Mart{\'{\i}}n-Navarro}, I., {La Barbera}, F., {Vazdekis}, A.,
  {Falc{\'o}n-Barroso}, J., \& {Ferreras}, I. 2015, \mnras, 447, 1033

\bibitem[{{Matteucci}(1994)}]{Matteucci1994}
{Matteucci}, F. 1994, \aap, 288, 57

\bibitem[{{Megeath} {et~al.}(2016){Megeath}, {Gutermuth}, {Muzerolle},
  {Kryukova}, {Hora}, {Allen}, {Flaherty}, {Hartmann}, {Myers}, {Pipher},
  {Stauffer}, {Young}, \& {Fazio}}]{Megeath2016}
{Megeath}, S.~T., {Gutermuth}, R., {Muzerolle}, J., {et~al.} 2016, \aj, 151, 5

\bibitem[{{Meurer} {et~al.}(2009){Meurer}, {Wong}, {Kim}, {Hanish}, {Werk},
  {Bland-Hawthorn}, {Zwaan}, {Koribalski}, {Thilker}, {Ferguson}, {Putman},
  {Knezek}, {Drinkwater}, {Hoopes}, {Meyer}, {Ryan-Weber}, \&
  {Staveley-Smith}}]{Meurer2009}
{Meurer}, G.~R., {Wong}, O.~I., {Kim}, J.~H., {et~al.} 2009, \apj, 695, 765

\bibitem[{{Mor} {et~al.}(2019){Mor}, {Robin}, {Figueras}, {Roca-F{\`a}brega},
  \& {Luri}}]{Mor2019}
{Mor}, R., {Robin}, A.~C., {Figueras}, F., {Roca-F{\`a}brega}, S., \& {Luri},
  X. 2019, \aap, 624, L1

\bibitem[{{Neijssel} {et~al.}(2019){Neijssel}, {Vigna-G{\'o}mez}, {Stevenson},
  {Barrett}, {Gaebel}, {Broekgaarden}, {de Mink}, {Sz{\'e}csi}, {Vinciguerra},
  \& {Mandel}}]{Neijssel19}
{Neijssel}, C.~J., {Vigna-G{\'o}mez}, A., {Stevenson}, S., {et~al.} 2019,
  \mnras, 490, 3740

\bibitem[{{Papadopoulos}(2010)}]{Papadopoulos2010}
{Papadopoulos}, P.~P. 2010, \apj, 720, 226

\bibitem[{{Pearson} {et~al.}(2018){Pearson}, {Wang}, {Hurley}, {Ma{\l}ek},
  {Buat}, {Burgarella}, {Farrah}, {Oliver}, {Smith}, \& {van der
  Tak}}]{Pearson18}
{Pearson}, W.~J., {Wang}, L., {Hurley}, P.~D., {et~al.} 2018, \aap, 615, A146

\bibitem[{{Pflamm-Altenburg} {et~al.}(2007){Pflamm-Altenburg}, {Weidner}, \&
  {Kroupa}}]{Pflamm-Altenburg07}
{Pflamm-Altenburg}, J., {Weidner}, C., \& {Kroupa}, P. 2007, \apj, 671, 1550

\bibitem[{{Pflamm-Altenburg} {et~al.}(2009){Pflamm-Altenburg}, {Weidner}, \&
  {Kroupa}}]{Pflamm2009}
{Pflamm-Altenburg}, J., {Weidner}, C., \& {Kroupa}, P. 2009, \mnras, 395, 394

\bibitem[{{Reddy} {et~al.}(2006){Reddy}, {Lambert}, \& {Allende
  Prieto}}]{Reddy06}
{Reddy}, B.~E., {Lambert}, D.~L., \& {Allende Prieto}, C. 2006, \mnras, 367,
  1329

\bibitem[{{Schneider} {et~al.}(2018){Schneider}, {Ram{\'{\i}}rez-Agudelo},
  {Tramper}, {Bestenlehner}, {Castro}, {Sana}, {Evans},
  {Sab{\'{\i}}n-Sanjuli{\'a}n}, {Sim{\'o}n-D{\'{\i}}az}, {Langer}, {Fossati},
  {Gr{\"a}fener}, {Crowther}, {de Mink}, {de Koter}, {Gieles}, {Herrero},
  {Izzard}, {Kalari}, {Klessen}, {Lennon}, {Mahy}, {Ma{\'{\i}}z Apell{\'a}niz},
  {Markova}, {van Loon}, {Vink}, \& {Walborn}}]{Schneider18}
{Schneider}, F.~R.~N., {Ram{\'{\i}}rez-Agudelo}, O.~H., {Tramper}, F., {et~al.}
  2018, \aap, 618, A73

\bibitem[{{Speagle} {et~al.}(2014){Speagle}, {Steinhardt}, {Capak}, \&
  {Silverman}}]{Speagle14}
{Speagle}, J.~S., {Steinhardt}, C.~L., {Capak}, P.~L., \& {Silverman}, J.~D.
  2014, \apjs, 214, 15

\bibitem[{{Spera} \& {Mapelli}(2017)}]{SperaMapelli17}
{Spera}, M. \& {Mapelli}, M. 2017, \mnras, 470, 4739

\bibitem[{{Steidel} {et~al.}(2016){Steidel}, {Strom}, {Pettini}, {Rudie},
  {Reddy}, \& {Trainor}}]{Steidel16}
{Steidel}, C.~C., {Strom}, A.~L., {Pettini}, M., {et~al.} 2016, \apj, 826, 159

\bibitem[{{Sukhbold} {et~al.}(2016){Sukhbold}, {Ertl}, {Woosley}, {Brown}, \&
  {Janka}}]{Sukhbold16}
{Sukhbold}, T., {Ertl}, T., {Woosley}, S.~E., {Brown}, J.~M., \& {Janka}, H.~T.
  2016, \apj, 821, 38

\bibitem[{{Sukhbold} \& {Woosley}(2014)}]{SukhboldWoosley14}
{Sukhbold}, T. \& {Woosley}, S.~E. 2014, \apj, 783, 10

\bibitem[{{Tolstoy} {et~al.}(2009){Tolstoy}, {Hill}, \& {Tosi}}]{Tolstoy09}
{Tolstoy}, E., {Hill}, V., \& {Tosi}, M. 2009, \araa, 47, 371

\bibitem[{{Tomczak} {et~al.}(2016){Tomczak}, {Quadri}, {Tran}, {Labb{\'e}},
  {Straatman}, {Papovich}, {Glazebrook}, {Allen}, {Brammer}, {Cowley},
  {Dickinson}, {Elbaz}, {Inami}, {Kacprzak}, {Morrison}, {Nanayakkara},
  {Persson}, {Rees}, {Salmon}, {Schreiber}, {Spitler}, \&
  {Whitaker}}]{Tomczak16}
{Tomczak}, A.~R., {Quadri}, R.~F., {Tran}, K.-V.~H., {et~al.} 2016, \apj, 817,
  118

\bibitem[{{Toonen} {et~al.}(2012){Toonen}, {Nelemans}, \& {Portegies
  Zwart}}]{Toonen12}
{Toonen}, S., {Nelemans}, G., \& {Portegies Zwart}, S. 2012, \aap, 546, A70

\bibitem[{{Vagnozzi} {et~al.}(2017){Vagnozzi}, {Freese}, \&
  {Zurbuchen}}]{Vagnozzi17}
{Vagnozzi}, S., {Freese}, K., \& {Zurbuchen}, T.~H. 2017, \apj, 839, 55

\bibitem[{{Vazdekis} {et~al.}(1997){Vazdekis}, {Peletier}, \&
  {Beckman}}]{Vazdekis1997}
{Vazdekis}, A., {Peletier}, R.~F., \& {Beckman}, J.~E. and.~{Casuso}, E. 1997,
  \apjs, 111, 203

\bibitem[{{Vink} \& {de Koter}(2005)}]{VinkDeKoter05}
{Vink}, J.~S. \& {de Koter}, A. 2005, \aap, 442, 587

\bibitem[{{Vink} {et~al.}(2001){Vink}, {de Koter}, \& {Lamers}}]{Vink01}
{Vink}, J.~S., {de Koter}, A., \& {Lamers}, H.~J.~G.~L.~M. 2001, \aap, 369, 574

\bibitem[{{Watts} {et~al.}(2018){Watts}, {Meurer}, {Lagos}, {Bruzzese},
  {Kroupa}, \& {Jerabkova}}]{Watts2018}
{Watts}, A.~B., {Meurer}, G.~R., {Lagos}, C.~D.~P., {et~al.} 2018, \mnras, 477,
  5554, in press

\bibitem[{{Weidner} {et~al.}(2013{\natexlab{a}}){Weidner}, {Ferreras},
  {Vazdekis}, \& {La Barbera}}]{Weidner13a}
{Weidner}, C., {Ferreras}, I., {Vazdekis}, A., \& {La Barbera}, F.
  2013{\natexlab{a}}, \mnras, 435, 2274

\bibitem[{{Weidner} \& {Kroupa}(2005)}]{WeidnerKroupa05}
{Weidner}, C. \& {Kroupa}, P. 2005, \apj, 625, 754

\bibitem[{{Weidner} {et~al.}(2013{\natexlab{b}}){Weidner}, {Kroupa},
  {Pflamm-Altenburg}, \& {Vazdekis}}]{Weidner13}
{Weidner}, C., {Kroupa}, P., {Pflamm-Altenburg}, J., \& {Vazdekis}, A.
  2013{\natexlab{b}}, \mnras, 436, 3309

\bibitem[{{Wheeler} {et~al.}(1989){Wheeler}, {Sneden}, \&
  {Truran}}]{WheelerSnedenTruran89}
{Wheeler}, J.~C., {Sneden}, C., \& {Truran}, Jr., J.~W. 1989, \araa, 27, 279

\bibitem[{{Whitaker} {et~al.}(2014){Whitaker}, {Franx}, {Leja}, {van Dokkum},
  {Henry}, {Skelton}, {Fumagalli}, {Momcheva}, {Brammer}, {Labb{\'e}},
  {Nelson}, \& {Rigby}}]{Whitaker14}
{Whitaker}, K.~E., {Franx}, M., {Leja}, J., {et~al.} 2014, \apj, 795, 104

\bibitem[{{Yan} {et~al.}(2017){Yan}, {Jerabkova}, \& {Kroupa}}]{Yan17}
{Yan}, Z., {Jerabkova}, T., \& {Kroupa}, P. 2017, \aap, 607, A126

\bibitem[{{Yan} {et~al.}(2019{\natexlab{a}}){Yan}, {Jerabkova}, \&
  {Kroupa}}]{Yan2019code}
{Yan}, Z., {Jerabkova}, T., \& {Kroupa}, P. 2019{\natexlab{a}}, {GalIMF:
  Galaxy-wide Initial Mass Function}

\bibitem[{{Yan} {et~al.}(2019{\natexlab{b}}){Yan}, {Jerabkova}, {Kroupa}, \&
  {Vazdekis}}]{Yan2019}
{Yan}, Z., {Jerabkova}, T., {Kroupa}, P., \& {Vazdekis}, A. 2019{\natexlab{b}},
  \aap, 629, A93

\bibitem[{{Zhang} \& {Zhao}(2005)}]{ZhangZhao05}
{Zhang}, H.~W. \& {Zhao}, G. 2005, \mnras, 364, 712

\bibitem[{{Zhang} {et~al.}(2018){Zhang}, {Romano}, {Ivison}, {Papadopoulos}, \&
  {Matteucci}}]{Zhang18}
{Zhang}, Z.-Y., {Romano}, D., {Ivison}, R.~J., {Papadopoulos}, P.~P., \&
  {Matteucci}, F. 2018, \nat, 558, 260

\bibitem[{{Zonoozi} {et~al.}(2019){Zonoozi}, {Mahani}, \&
  {Kroupa}}]{Zonoozi2019}
{Zonoozi}, A.~H., {Mahani}, H., \& {Kroupa}, P. 2019, \mnras, 483, 46

\end{thebibliography}
\end{document}